\documentclass[onecolumn]{IEEEtran}
\usepackage{amsthm}
\usepackage{times,amssymb,amsmath,amsfonts,float,nicefrac,color,bbm,mathrsfs,caption,float}
\usepackage{algorithm,enumerate,multirow,caption,tikz,graphicx}
\usepackage[mathscr]{eucal}
\usepackage{sidecap}
\usepackage{algpseudocode}
\usepackage{verbatim}
\usepackage{epstopdf}
\usepackage{textcomp}
\usetikzlibrary{shapes,arrows}
\usepackage{stmaryrd}
\usepackage{mathabx}
\usepackage[noadjust]{cite}
\usepackage{booktabs}
\usepackage[normalem]{ulem}

\usepackage[top=1.3in,bottom=1.3in,left=1.5in,right=1.5in]{geometry}
\allowdisplaybreaks

\newcommand{\RN}[1]{%
  \textup{\expandafter{\romannumeral#1}}%
}

\newcommand\remove[1]{}

\allowdisplaybreaks

\def\mathbi#1{{\textbf{\textit #1}}}

\newtheorem{theorem}{Theorem}
\newtheorem{definition}{Definition}
\newtheorem{proposition}{Proposition}

\newtheorem{claim}{Claim}
\newtheorem{lemma}{Lemma}

\newtheorem{example}{Example}
\newtheorem{remark}{Remark}

\newcommand{\ff}{{\mathbb F}}

\newcommand{\cC}{\mathcal{C}}
\newcommand{\cF}{\mathcal{F}}

\newcommand{\cJ}{\mathcal{J}}

\newcommand{\cP}{\mathcal{P}}\newcommand{\cR}{\mathcal{R}}
\newcommand{\cS}{\mathcal{S}}

\DeclareMathOperator{\trace}{tr}

\DeclareMathOperator{\lcm}{lcm}

\DeclareMathOperator{\Diag}{Diag}
\DeclareMathOperator{\spun}{Span}

\begin{document}
\title{The repair problem for Reed-Solomon codes: Optimal repair of single and multiple erasures, asymptotically optimal node size}

\author{\IEEEauthorblockN{Itzhak Tamo} \hspace*{1in}
\and \IEEEauthorblockN{Min Ye} \hspace*{1in}
\and \IEEEauthorblockN{Alexander Barg}}

\maketitle
{\renewcommand{\thefootnote}{}\footnotetext{

\vspace{-.2in}
 
\noindent\rule{1.5in}{.4pt}

{The results of Sections \ref{sect:single-node} and \ref{Sect:lb} were presented at the 58th IEEE Symposium on the Foundations of Computer Science (FOCS), October 15-17, 2017, Berkeley, CA \cite{Tamo17RS}, and the result of Section~\ref{RS} was presented at the 2016 IEEE International Symposium on Information Theory, July 2016, Barcelona, Spain \cite{Ye16b}.

I. Tamo is with Department of EE-Systems, Tel Aviv University, Tel Aviv,
Israel. Email: zactamo@gmail.com.
His research is supported by ISF grant no.~1030/15 and the NSF-BSF grant no.~2015814.

M. Ye was with Department of ECE and ISR, University of Maryland, College Park, MD 20742. Email: yeemmi@gmail.com
His research was supported by NSF grant CCF1422955.

A. Barg is with Dept. of ECE and ISR, University of Maryland, College Park, MD 20742 and also with IITP, Russian Academy of Sciences, 127051 Moscow, Russia. Email: abarg@umd.edu. His research was supported by NSF grants CCF1422955 and CCF1618603.}}
}
\renewcommand{\thefootnote}{\arabic{footnote}}
\setcounter{footnote}{0}

\begin{abstract}
The repair problem in distributed storage addresses recovery of the data encoded using an erasure code, for instance, a Reed-Solomon (RS) code. We consider the problem of repairing a single node or multiple nodes in RS-coded storage systems using the smallest possible amount of inter-nodal communication. According to the cut-set bound, communication cost of repairing $h\ge 1$ failed nodes for an $(n,k=n-r)$ MDS code using $d$ helper nodes is at least $dhl/(d+h-k),$ where $l$ is the size of the node. Guruswami and Wootters (2016) initiated the study of efficient repair of RS codes, showing that they can be repaired using a smaller bandwidth than under the trivial approach. At the same time, their work as well as follow-up papers stopped short of constructing RS codes (or any scalar MDS codes) that meet the cut-set bound with equality.

In this paper we construct families of RS codes that achieve the cutset bound for repair of one or several nodes. In the single-node case, we present RS codes of length $n$ over the field $\ff_{q^l},
l=\exp((1+o(1))n\log n)$ that meet the cut-set bound. We also prove an almost matching lower bound on $l$, showing that super-exponential scaling is both necessary and sufficient for scalar MDS codes to achieve the cut-set bound using linear repair schemes. For the case of multiple nodes, we construct a family of RS codes that achieve the cut-set bound universally for the repair of any $h=2,3,\dots$ failed nodes from any subset of $d$ helper nodes, $k\le d\le n-h.$ For a fixed number of parities $r$ the node size of the constructed codes is close to the smallest possible node size for codes with such properties. 
\end{abstract}

\clearpage
\tableofcontents

\section{Introduction}\label{section:repairing RS}

\subsection{Minimum Storage Regenerating codes and optimal repair bandwidth}

The problem considered in this paper is motivated by the distributed nature of the system wherein the coded data is distributed across a large number of physical storage nodes. When some storage nodes fail,
the repair task performed by the system relies on communication between individual nodes, which introduces new challenges in the code design. In particular, a new parameter that has a bearing on the overall efficiency of the system is the {\em repair bandwidth}, i.e., the amount of data communicated between the nodes in the process of repairing failed nodes. 

Modern large-scale distributed storage systems rely on information encoding using Maximum Distance Separable (MDS) codes since they provide the optimal tradeoff between failure tolerance and storage overhead. 
To encode information with an MDS code, we represent data chunks as elements of a finite field. More specifically, we divide the original file into $k$ information blocks and view each block as a single element of a finite field $F$ or a vector over $F$. We encode the data by adding $r=n-k$ parity blocks (field symbols or vectors) and distribute the resulting $n$ blocks across $n$ storage nodes. The MDS property ensures that the original file can be recovered from the content stored on any $k$ nodes. In this paper we deal only with linear codes, so the parity blocks are formed as linear combinations of the information blocks over $F.$ We use the notation $(n,k)$ to refer to the length and dimension of a linear code.


Before proceeding further, we make a brief remark on the terminology used in the literature devoted to erasure correcting codes for
distributed storage. The coordinates of the codeword are assumed to be stored on different nodes, and by extension are themselves 
referred to as nodes. 
We assume that the data is encoded with a
code $\cC$ over a finite field $F$ wherein each coordinate of the codeword is either an element of $F$ or an $l$-dimensional vector
over $F$, where $l> 1.$ 
The latter construction, termed {\em array codes} turns out to be better suited to the needs of the repair problem,
as will be apparent in the later part of this section.  To repair a failed node, the system needs to download
the contents from some other nodes ({\em helper nodes}) of the codeword to the processor, and the total amount of the downloaded data is 
called the {\em repair bandwidth}. Coding solutions that support efficient repair are called {\em regenerating codes}, and they
have been a focal point of current research in coding theory following their introduction in {\sc Dimakis et al.}~\cite{Dimakis10}.

One traditional solution to recover a single node failure in an MDS-coded system is to download the content stored on any $k$ nodes. The 
MDS property guarantees that we can recover the whole file, so we can also recover any single node failure. However, this method is far 
from efficient in the sense that the repair bandwidth that it requires is much larger than {is needed} for the repair of a single 
node. Indeed, by a rather counter-intuitive result of \cite{Dimakis10} it is possible to save on the repair bandwidth by
contacting $d>k$ helper nodes, and the maximum savings are attained when $d$ is the largest possible value, namely $d=n-1$. 

More specifically, suppose that an $(n,k)$ MDS-coded system attempts to repair a failed node by connecting to $d$ helper nodes. In this case, as shown in \cite{Dimakis10}, the total amount of information that needs to be downloaded
to complete the repair task is at least $dl/(d+1-k),$ where $l$ is the size of each node. This lower bound on the repair bandwidth is called the {\em cut-set bound} since it is obtained from the cut-set bound in network information theory \cite{ElGamal81}.
Given $k< d\le n-1,$ an $(n,k)$ MDS code achieving the cut-set bound for the repair of any single failed node from any $d$ helper 
nodes is called an $(n,k)$ {\em minimum storage regenerating} (MSR) code with {\em repair degree} $d$ \cite{Dimakis10}.

The definition of MSR codes, given above in an informal way, will be formalized for a particular subclass of codes known as {\em MDS array codes}. 
An $(n,k)$ MDS array code $\cC$ with {\em sub-packetization} $l$ over a finite field $F$ is formed of
$k$ information nodes and $r=n-k$ parity nodes, where every node is a column vector of length $l$ over $F$ (so $\dim_F (\cC)= kl$).
The MDS property requires that any $k$ nodes of $\cC$ suffice to recover the remaining $r$ nodes of the codeword. Array codes are also called {\em vector codes}, while code families more common to coding theory  
 (such as Reed-Solomon (RS) codes and others) are called {\em scalar codes} in the literature. Clearly, scalar codes correspond to the case $l=1$ of the above definition.

%

Throughout the paper we use the notation $[n]:=\{1,2,\dots,n\}.$
Consider an $(n,k,l)$ array code $\cC$ over a finite field $F$. We write a codeword of $\cC$ as $c=(c_1,\dots,c_n)$,
where $c_i=(c_{i,0},c_{i,1},\dots,c_{i,l-1})^T\in F^l, i=1,\dots, n$. 
A node $c_i,i\in [n]$ can be repaired from a subset of $d\ge k$ helper nodes $\{c_j:j\in\cR\},\cR\subseteq[n]\backslash \{i\},$
by downloading $\beta_i(\cR)$ symbols of $F$ if there are
numbers $\beta_{ij}, j\in\cR$,
functions $f_{ij}: F^l\to F^{\beta_{ij}}, j\in\cR,$
and a function $g_i: F^{\sum_{j\in\cR}\beta_{ij}}\to F^l$
such that
   $$
   c_i=g_i(\{f_{ij}(c_{j}),j\in\cR \})
\text{~for all~} c=(c_1,\dots,c_n)\in \cC
   $$
   and
   $$
\sum_{j\in\cR}\beta_{ij}=\beta_i(\cR).
   $$
This definition extends straightforwardly to the repair of a subset of failed nodes $\{c_i:i\in\cF\},\cF\subseteq [n]$ from a subset of helper nodes $\{c_j:j\in\cR\},\cR\subseteq[n]\backslash \cF$.
We note that the symbols downloaded to repair the failed node(s) can be some functions of the contents  of the helper
nodes $c_j,j\in{\cR}$.

\begin{definition}[Repair bandwidth]
Let $\cC$ be an $(n,k,l)$ {MDS} array code over a finite field $F$ and let $c=(c_1,\dots,c_n)\in \cC$ be a codeword.
Given two disjoint subsets ${\cF},{\cR}\subseteq[n]$ such that $|{\cF}|\le r$ and $|{\cR}|\ge k,$ we define $N(\cC,{\cF},{\cR})$ as the smallest number of symbols of $F$ one needs to download from the helper nodes $\{c_i:i\in{\cR}\}$ in order to recover the failed (erased) 
nodes $\{c_i:i\in{\cF}\}.$  The \emph{$(h,d)$-repair bandwidth} of the code $\cC$ equals 
    \begin{equation}\label{eq:beta}
    \beta(h,d):=\max_{|{\cF}|=h,|{\cR}|=d, {\cF}\bigcap{\cR}=\emptyset} N(\cC,{\cF},{\cR}).
    \end{equation}
\end{definition}

The following basic result sets a benchmark for the minimum repair bandwidth.
\begin{theorem}[Cut-set bound, {\sc Dimakis et al.}~\cite{Dimakis10}, {\sc Cadambe et al.} \cite{Cadambe13}] \label{def:csb}
Let $\cC$ be an $(n,k,l)$ MDS array code. For any two disjoint subsets ${\cF},{\cR}\subseteq[n]$ such that $|{\cF}|\le r$ and $|{\cR}|\ge k,$ we have the following inequality:
\begin{equation}\label{eq:cutset}
N(\cC,{\cF},{\cR})\ge \frac{|{\cF}||{\cR}|l}{|{\cF}|+|{\cR}|-k}.
\end{equation}
\end{theorem}

\begin{definition}
We say that an $(n,k,l)$ MDS code $\cC$ has the \emph{$(h,d)$-optimal repair} property if the $(h,d)$-repair bandwidth of $\cC$ (see \eqref{eq:beta})
equals 
   \begin{equation}\label{eq:cs1}
   \beta(h,d)=\frac{hdl}{h+d-k},
   \end{equation}
meeting the lower bound in \eqref{eq:cutset} with equality.
 \end{definition}
 
Another important parameter is the value of sub-packetization $l$.  Due to the limited storage capacity of each node, we would like $l$ to be as small as possible. However, it is shown in \cite{Goparaju14} that for an $(n,k,d=n-1,l)$ MSR array code,  $l\ge \exp({\sqrt{k/(2r-1)}})$  (i.e., $l$ is exponential in $n$ for fixed $r$ and growing $n$). 
 
Several constructions of MDS array codes with optimal repair property are available in the literature. For the case of low 
code rate where $k\le n/2$, optimal-repair codes were constructed by {\sc Rashmi, Shah, and Kumar} \cite{Rashmi11}. Constructions that have no
rate limitations were proposed by
{\sc Tamo et al.}~\cite{Tamo13}, {\sc Ye and Barg} \cite{Ye16,Ye16a}, {\sc Goparaju et al.}~\cite{Goparaju17}, {\sc Raviv et al.}~\cite{Raviv17}. In particular, \cite{Ye16} gave explicit constructions of MDS array codes with the universal $(h,d)$-optimal repair property for all $h\le r$ and all $k\le d\le n-h$ simultaneously. In other words, the codes in \cite{Ye16} can repair any number of erasures $h$ from any set of $d$ helper nodes with the repair bandwidth achieving the cut-set bound \eqref{eq:cs1}.

As a final remark, note that two models of repairing multiple node failures are commonly used in the literature. The prevalent one is the {\em centralized model}, where a single repair center is responsible for the repair of all failed nodes \cite{Cadambe13,Ye16,Rawat16a,Wang17,Zorgui17}. The other one is the {\em cooperative model}, where the failed nodes may cooperate but are distinct, and the amount of data communicated between the failed nodes is also included in the repair bandwidth \cite{Kermarrec11,Shum13,Li14}. The version of the cut-set bound in \eqref{eq:cutset} is derived under the centralized model; see \cite{Cadambe13}. Moreover, it is shown in \cite{Shum13} that \eqref{eq:cutset} is not achievable under the cooperative model (they also derive a version of the bound \eqref{eq:cutset} that applies in the cooperative case). 
Optimal-repair MDS array codes for the cooperative case were recently constructed in \cite{YeBarg18}.
In this paper we only consider the centralized model.


 \subsection{Repair schemes for scalar linear MDS codes}
While there has been much research into constructions and properties of MSR codes specifically designed for the repair task, it is
also of interest to study the repair bandwidth of general families of MDS codes, for instance, RS codes.
In \cite{Shanmugam14}, {\sc Shanmugam et al.} proposed a framework for studying the repair bandwidth of a scalar linear $(n,k)$ MDS code $\cC$ 
over some finite field $E$ (called symbol field below). The idea of \cite{Shanmugam14} is to ``vectorize'' the code construction by considering $\cC$ 
as an array code over some subfield $F$ of $E$. This approach provides a bridge between RS codes and MDS array codes, wherein 
the extension degree $l:=[E:F]$ can be viewed as the value of sub-packetization. The code $\cC$ is viewed as an $(n,k)$ MDS array code with sub-packetization $l$, and the repair bandwidth is
defined exactly in the same way as above. The cut-set bound \eqref{eq:cutset} {and the definition of MSR codes} also apply to this setup.

In this paper we study repair of RS codes, focusing on linear repair schemes, i.e., we assume that the repair operations
are linear over the field  $F.$

In {\sc Guruswami and Wootters}~\cite{Guruswami16}, there is one more restriction on the parameters of the RS codes, namely they achieve the smallest possible
repair bandwidth only if the number of parities is of the form $r=q^s,(l-s)|l.$ In \cite{Dau17}, {\sc Dau and Milenkovic} generalized the scheme in \cite{Guruswami16} and extended their
results to all values of $s=1,\dots,l-1$. The repair bandwidth attained in \cite{Dau17} is $(n-1)(l-s)$ symbols of $F$
 for $r\ge q^s$, and is the smallest possible whenever $r$ is a power of $q.$ 
 Several works also extended the framework of \cite{Guruswami16} to the repair of more than one erasure (node failure) for RS codes,
see {\sc Dau et al.}~\cite{Dau16}, {\sc Mardia et al.}~\cite{Bartan17}. At the same time, \cite{Guruswami16} as well as follow-up papers stopped short of constructing RS codes (or any scalar MDS codes) that meet the cut-set bound \eqref{eq:cs1} with equality (no matter for repairing single erasure or multiple erasures).
All the previous papers (apart from {\sc Ye and Barg}~\cite{Ye16b}) focused on small sub-packetization regime, and the repair bandwidth of their constructions is rather far from the cut-set bound.

To summarize the earlier work, constructions of RS codes (or any scalar MDS codes) that meet the cut-set bound have as yet been unknown, so the existence question of such codes has been an open problem. In this paper, we resolve this problem in the affirmative, presenting such a construction. We also prove a lower bound on the sub-packetization of scalar linear MDS codes that attain the cut-set bound with a linear repair scheme, showing that there is a penalty for the scalar case compared to MDS array codes.

\subsection{Our Results}\label{sect:results}

\subsubsection
{Explicit constructions of $(1,d)$ optimal-repair RS codes} Given any $n,k$ and $d, k\le d\le n-1$, we construct an $(n,k)$ RS code over the field $E=\ff_{q^l}$ that achieves the cut-set bound \eqref{eq:cutset} when repairing {\em any} single failed node from {\em any} $d$ helper nodes. As above, we view RS codes over $E$ as vector codes over the subfield $F=\ff_q$. 
The main novelty in our construction is the choice of the evaluation points for the code in such a way their degrees distinct primes. For the actual repair we rely on the linear scheme proposed in \cite{Guruswami16} presented below in Sec.~\ref{sect:GW} (this is essentially the only possible linear repair approach).

The value of sub-packetization $l$ of our construction equals $s$ times the product of the first $n$ 
distinct primes in an arithmetic progression, 
  \begin{equation}\label{eq:ell1}
   l= s \biggl(\prod_{\substack{i=1\\[.02in]p_i\equiv 1\text{ mod } s}}^n p_i\biggr),
  \end{equation}
  where $s:= d+1-k.$ To quantify the behavior of \eqref{eq:ell1} for large $n$, note that this product is a well-studied function in number theory, related to a classical arithmetic function $\psi(n,s,a)$  (which is essentially the sum of logarithms of the primes). The prime number theorem in arithmetic progressions (for instance, \cite[p.121]{IK04}) yields asymptotic estimates for $l$. In particular, for fixed $s$ and large $n$, we have $l= e^{(1+o(1)) n\log n}.$ 

In contrast, for the case $d=n-1$ (i.e., $s=r=n-k$), there exist MSR array codes that attain sub-packetization $l=r^{\lceil n/(r+1) \rceil}$ \cite{Wang16},
which is the smallest known value among MSR codes\footnote{ The construction of \cite{Wang16} achieves the cut-set bound only for repair of systematic nodes, and gives $l=r^{\lceil k/(r+1) \rceil}$. Using the approach of \cite{Ye16}, it is possible to modify the construction of \cite{Wang16} and to obtain an MSR code with $l=r^{\lceil n/(r+1) \rceil}$.}.  So although this  distinct prime structure allows us to achieve the cut-set bound, it makes us pay a penalty on the sub-packetization. 

\vspace*{.1in}\subsubsection {A lower bound on the sub-packetization of scalar MDS codes achieving the cut-set bound} 
Surprisingly, we also show that the distinct prime structure discussed above is necessary for any scalar linear MDS code (not just the RS codes) to achieve the cut-set bound under linear repair. Namely, given $d$ such that $k+1\le d\le n-1,$ we prove that for any $(n,k)$ scalar linear MSR code with repair degree
$d,$ 
 the sub-packetization $l$ is bounded below by  $l \geq \prod_{i=1}^{k-1}p_i$, where $p_i$ is the $i$-th smallest prime.  By the Prime Number 
Theorem \cite{IK04}, we obtain the lower asymptotic bound on $l$ of the form
$
l\ge e^{(1+o(1))k \log k}.
$

In summary, we obtain the following results for the smallest possible sub-packetization of scalar linear MDS codes, including the RS codes, whose repair bandwidth 
achieves the cut-set bound.

\begin{theorem}\label{thm:main} Let $\cC$ be an $(n,k=n-r)$ scalar linear MDS code over the field $E=\ff_{q^l},$ 
and let $d$ be an integer satisfying $k+1\le d\le n-1.$ Suppose that for any single failed node of $\cC$ and any $d$ helper nodes there
is a linear repair scheme over $\ff_q$ that uses the bandwidth $dl/(d+1-k)$ symbols of $\ff_q$, i.e., it achieves the cut-set bound \eqref{eq:cutset}. For a fixed $s=d+1-k$ and $n,k\to\infty$ the following bounds on the smallest possible sub-packetization hold true:
  \begin{equation}\label{eq:main} 
  e^{(1+o(1))k \log k}\le l\le e^{(1+o(1)) n\log n}.
  \end{equation}
For large $s$, we have $l\le 
s \prod\limits_{\substack{i: p_i\equiv 1\,\text{\rm mod}\, s}}^n p_i,$
where the product goes over the first $n$ distinct primes in the arithmetic progression.
\end{theorem}
\begin{remark} {\rm The bound on $l$ can be made more explicit 
even for large $s$, and the answer depends on whether we accept the Generalized Riemann Hypothesis (if yes, we can still claim the bound 
$l\le \exp((1+o(1))n\log n)$).} \end{remark}

Theorem \ref{thm:main} will follow from Theorems \ref{thm1} and \ref{thm2} proved below in the paper.
\vspace*{.1in}

\subsubsection{Repairing multiple erasures: $(h,d)$-optimal RS codes for all admissible parameters}
Developing the ideas in Part (1), we also construct a family of RS codes that support optimal repair multiple nodes from any subset of 
helper nodes. Our results in this part are formulated as follows.
\begin{theorem}\label{thm:multiple} $(i)$ For any $k<n$ there exists an explicitly constructible family of $(n,k)$ RS codes over a suitably chosen
finite field $\ff_{q^l}$ with the $(2,d)$ optimal repair property and sub-packetization
   \begin{equation}\label{eq:ell2}
    l=(d-k+1)(d-k+2)e^{(1+o(1))n\log n}.
    \end{equation}
(ii) There exists an explicitly constructible family of $(n,k)$ RS codes over a suitably chosen
finite field $\ff_{q^l}$ with the universal $(h,d)$ optimal repair property for all $h\le r$ and $k\le d\le n-h$ simultaneously,
where
    \begin{equation}\label{eq:elld}
    l=r!\,e^{(1+o(1))n\log n}.
    \end{equation}
\end{theorem}    
The statements of this theorem are made more precise in Theorems \ref{thm:2err} and \ref{thm:ue} below.
According to the lower bound in \eqref{eq:main}, when the code rate $k/n$ is close to $1$, the sub-packetization value of our codes is close to the optimal value among all scalar linear MDS codes with the optimal repair property.

\vspace*{.1in}\subsubsection{RS codes with asymptotically optimal $(1,n-1)$ repair and $l=r^n$} We also point out that the values of $l$ for single-node repair can be reduced if instead of exact optimality we achieve asymptotic optimality of the repair bandwidth in the large $n$ regime.
Specifically, the following is true.
  \begin{theorem}\label{thm:as}
  There exists an explicitly constructible family of $(n,k)$ RS codes over $\ff_{q^l},l=r^n$ with repair bandwidth at most $l\frac{n+1}{n-k}.$
  \end{theorem}
This result, which is a direct development of the work in \cite{Guruswami16}, is formalized in Theorem \ref{thm:powers}.

\subsection{Discussion: Array codes and scalar codes} The lower bound in \eqref{eq:main} is 
much larger than the sub-packetization of many known MSR array code constructions (for instance, there are MSR array codes with 
$l=r^{\lceil n/r\rceil}$ \cite{Ye16a,Sasid16}, and an impossibility result in \cite{BalajiKumar2017} shows that this construction is optimal 
in terms of $l$).
To make clearer the comparison between the repair parameters of scalar codes and array codes, we summarize the tradeoff between the repair bandwidth and the sub-packetization of some known MDS code constructions in Table~\ref{table1}.
We list only results considering the repair of a single node from all the remaining $n-1$ helper nodes. Moreover, in the table we limit ourselves to explicit code constructions, and do not list multiple existence results that appeared in recent years.

\begin{table}
\captionof{table}{Tradeoff between repair bandwidth and sub-packetization}
\begin{center}{\normalsize \begin{tabular}{| c | c | c | c |} 
\hline Code construction
  & Repair bandwidth &  sub-packetization  & achieving cut-set bound \\ \hline\multicolumn{4}{c}{Array codes}\\\hline
\begin{tabular}[c]{@{}c@{}c@{}} $(n,k=n-r,n-1,l)$\\ MSR array codes for \\ $2k\le (n+1)$, \cite{Rashmi11} 
\end{tabular} & $\frac{(n-1)l}{r}$  & $l=r$ & Yes  \\ \hline
\begin{tabular}[c]{@{}c@{}c@{}} $(n,k,n-1,l)$\\ MSR array codes \\ (a modification of \cite{Wang16}) 
\end{tabular} & $\frac{(n-1)l}{r}$  & $l=r^{\lceil n/(r+1) \rceil}$ & Yes  \\ \hline
\begin{tabular}[c]{@{}c@{}c@{}} $(n,k,n-1,l)$  MSR \\ array codes \cite{Ye16a} 

\end{tabular} & $\frac{(n-1)l}{r}$  & $l=r^{\lceil n/r\rceil}$ & Yes  \\ \hline
\begin{tabular}[c]{@{}c@{}c@{}} $(n,k)$ MDS \\ array codes with design \\ parameter $t\ge1$
\cite{Guruswami17}
\end{tabular} & $(1+\frac{1}{t})\frac{(n-1)l}{r}$  & $l=r^t$ & No  \\  \hline \multicolumn{4}{c}{Scalar codes}\\ \hline
\begin{tabular}[c]{@{}c@{}} $(n,k)$ RS code  \cite{Ye16b}
\end{tabular} & $<\frac{(n+1)l}{r}$  & $l=r^n$ & No  \\  \hline
\begin{tabular}[c]{@{}c@{}} $(n,k)$ RS code  \cite{Guruswami16}
\end{tabular} & $n-1$  & $l=\log_{n/r} n$ & No  \\   \hline
\begin{tabular}[c]{@{}c@{}} $(n,k)$ RS code  \cite{Dau17}\\  
\end{tabular} &  $(n-1)l(1-\log_nr)$ & $\log_q n$ & No  \\   \hline
\begin{tabular}[c]{@{}c@{}} $(n,k)$ RS code\\  (this paper)
\end{tabular} & $\frac{(n-1)l}{r}$  & $l\approx n^{ n}$ & Yes  \\ 
\hline
\end{tabular}}
\end{center}\label{table1}
\end{table}

As already mentioned, the constructions of \cite{Guruswami16,Dau17} have optimal repair bandwidth among all the RS codes with the same sub-packetization value as in these papers\footnote{Expressing the sub-packetization of the construction in \cite{Dau17} via $n$ and $r$ is
difficult. The precise form of the result in \cite{Dau17} is as follows: for every $s<l$ and $r\ge q^s,$ the authors construct repair schemes of RS codes of length $n=q^l$
with repair bandwidth $(n-1)(l-s).$ Moreover, if $r=q^s,$ then the schemes proposed in \cite{Dau17} achieve the smallest possible repair bandwidth for codes with these parameters.}.
At the same time, these values are too small for the constructions of \cite{Guruswami16,Dau17} to achieve the cut-set bound.
From the first three rows of the table one can clearly see that the achievable sub-packetization values for MSR array codes are much smaller than the lower bound for scalar linear MSR codes derived in this paper. This is to be expected since for array codes we only require the code to be linear over the ``repair field,'' i.e., $F$, and not the symbol field $E$ as in the case of scalar codes. 

\subsection{Organization of the paper}
Our results are presented in Sections \ref{sect:single-node}--\ref{RS}.
Namely, in Sec.~\ref{sect:warmup}, we present a {simple construction} of RS codes that achieve the cut-set bound for repair of 
a subset of the nodes (not necessarily systematic). This construction is inferior to the more involved construction of Sec.~\ref{Sect:cons}, but simple to follow, and already contains some
of the main ideas of the general case, so we include it as a warm-up for the later results. In Sec.~\ref{Sect:cons}, we
present our main construction of RS codes that achieve the cut-set bound for the repair of any single node, proving the upper estimate in \eqref{eq:main}. In Sec.~\ref{Sect:lb}, we prove the lower bound on the sub-packetization of scalar linear MSR codes, finishing the proof of \eqref{eq:main}. The results of this part of the paper were presented earlier at FOCS'17 and published in \cite{Tamo17RS}.

The second part is devoted to a construction of RS codes with optimal repair of multiple erasures. In Sec.~\ref{sect:warmup-multiple}
we present the case of $h=2$ failed nodes, which captures the ideas of the transition from the single-node case to several nodes.
These ideas are developed in Sec.~\ref{Sect:uee} where we present a family of RS codes with universally optimal repair of any
$h\le n-k$ failed codes from any $k\le d\le n-h$ helper nodes, proving Theorem~\ref{thm:multiple}. The presentation is rather
technical, which is why we added Sec.~\ref{sect:warmup-multiple} to make it more accessible.

Finally, in Sec.~\ref{RS} we present a simple construction of RS codes that {\em asymptotically} achieve the optimal 
bandwidth, using sub-packetization smaller that in the finite-length constructions above ($r^n$ compared to about $n^n$).
This construction was presented earlier at ISIT'16 and published as a part of the extended abstract \cite{Ye16b}.

\section{Some definitions}\label{sect:definitions} Let us first recall some basic concepts that will be used throughout the paper.

\begin{definition}[Dual code] Let $\cC$ be a linear code of length $n$ over a finite field $F$. 
The dual code of $\cC$ is the linear subspace of $F^n$ defined by
$$
\cC^{\perp}=\big\{x=(x_1,\dots,x_n) \in F^n \big|\sum_{i=1}^n x_i c_i = 0 \quad \forall c=(c_1,\dots c_n)\in\cC \big\}.
$$
\end{definition}

\begin{definition}
A \emph{Generalized Reed-Solomon code} $\text{\rm GRS}_F(n,k,\Omega,v)\subseteq F^n$ of dimension $k$ over a field $F$ 
with evaluation points $\Omega=\{\omega_1,\omega_2,\dots,\omega_n\}\subseteq F$  is the set of vectors
\begin{align*}
\{(v_1f(\omega_1),\dots,v_nf(\omega_n))\in F^n:f\in F[x], \deg f\le k-1\}
\end{align*}
where $v=(v_1,\dots,v_n)\in (F^\ast)^n$ are some nonzero elements. If $v=(1,\dots,1),$ then the GRS code is called
a Reed-Solomon code and is denoted as $\text{\rm RS}_F(n,k,\Omega)$.

It is well known \cite[p.304]{Macwilliams77} that 
   \begin{equation}\label{eq:grs}
    (\text{\rm RS}_F(n,k,\Omega))^\bot=\text{\rm GRS}_F(n,n-k,\Omega,v)
  \end{equation}
where $v_i=\prod_{j\ne i}(\omega_i-\omega_j)^{-1}, i=1,\dots,n$. (The dual of an RS code is a GRS code.)  
\end{definition}   

Let $F$ be a finite field and let $E$ be the extension of $F$ of degree
$t.$ The trace function $\trace_{E/F}:E\to F$ is defined by
  $$
   \trace_{E/F}(x):=x+x^{|F|}+x^{|F|^2}+\dots+x^{|F|^{t-1}}.
   $$
The trace has the following \emph{transitivity property}: let $K$ be a finite algebraic extension of $E$, then for all $a\in K,$
   \begin{equation}\label{eq:trans}
    \trace_{K/F}(a)=\trace_{E/F}(\trace_{K/E}(a)).
   \end{equation}

\section{The linear repair scheme of Guruswami and Wootters \cite{Guruswami16}}\label{sect:GW}
 Suppose the symbol field of the code $\cC=\text{RS}(n,k,\Lambda)$ is $E$ and we want to repair it over the base field $F\subseteq E.$ More precisely, if a single codeword symbol is erased, we will recover this symbol by download sub-symbols of the base field $F$ from the surviving nodes. In order to make the repair scheme $F$-linear, \cite{Guruswami16} uses 
$F$-linear transforms $L_{\gamma}:E\to F$ given by the \emph{trace functionals} $L_{\gamma}(\beta)=\text{tr}(\gamma\beta).$ 

Let $\{\zeta_1,\dots,\zeta_l\}$ be a basis for $E$ over $F,$ and let $\{\mu_1,\dots,\mu_l\}$ be its dual (trace-orthogonal) basis,
namely $\trace_{E/F}(\zeta_i\mu_j)=\delta_{ij}.$ The coefficients of the expansion of an element $\beta\in E$ in the basis
$(\mu_i)$ are given by $\trace(\zeta_i\beta),$ so that 
   \begin{equation}\label{eq:expansion}
\beta=\sum_{i=1}^l (\trace(\zeta_i\beta)\mu_i).
   \end{equation}

Let $\cC^{\perp}$ be the dual code of $\cC=\text{RS}(n,k,\Lambda).$ 
Suppose that the codeword symbol $c_i$ in a codeword $c=(c_1,\dots,c_n)\in\cC$ is erased. We can find $l$ codewords 
$\{c_j^\bot=(c_{j,1}^{\perp},\dots,c_{j,n}^{\perp})\}_{j=1}^l$ in $\cC^{\perp}$ such that 
$\{c_{1,i}^{\perp},\dots,c_{l,i}^{\perp}\}$ is a basis of $E$ over $F.$ By the observation above, knowing the values of 
$\{\text{tr}(c_{j,i}^{\perp}c_i)\}_{j=1}^l$ suffices to recover the erased symbol $c_i.$ Since the trace is an $F$-linear transformation, we have
$$
\text{tr}(c_{j,i}^{\perp}c_i)=-\sum_{t\neq i}\text{tr}(c_{j,t}^{\perp}c_t) \text{ for all } j\in[l].
$$
Thus knowing the values of $\{\{\text{tr}(c_{j,t}^{\perp}c_t)\}_{j\in[l]}\}_{t\in[n],t\neq i}$ suffices to recover $c_i.$ Let 
$B_t$ be a maximal linearly independent subset of the set $\{c_{j,t}^{\perp}\}_{j\in[l]}$ over $F.$ Again due to the $F$-linearity of
the trace function, $\{\text{tr}(c_{j,t}^{\perp}c_t)\}_{j\in[l]}$ can be calculated from 
$\{\text{tr}(\beta c_t)\}_{\beta\in B_t}.$ Consequently, $c_i$ can be recovered from $\{\{\text{tr}(\beta c_t)\}_{\beta\in 
B_t}\}_{t\in[n],t\neq i}.$ The total number of sub-symbols in $F$ we need to download from the surviving nodes to recover $c_i$ is 
$\sum_{t\in[n],t\neq i} {\dim}_F(\{c_{j,t}^{\perp}\}_{j\in[l]}).$ 

\vspace*{.05in} We conclude that to efficiently recover $c_i,$ we need to find $l$ codewords in $\cC^{\perp}$ 
that minimize the quantity $\sum_{t\in[n],t\neq i} {\dim}_F(\{c_{j,t}^{\perp}\}_{j\in[l]})$ under the condition 
that $\{c_{1,i}^{\perp},\dots,c_{l,i}^{\perp}\}$ is a basis for $E$ over $F.$ 

As already remarked, $\cC^{\perp}=\text{GRS}(n,n-k,\Lambda,v)$ for some nonzero coefficients
$v=(v_1,\dots,v_n)\in E^n.$ 
Choosing a codeword from $\cC^{\perp}=\text{GRS}(n,n-k,\Lambda,v)$ is equivalent to choosing a polynomial with degree less than $n-k.$ Suppose $\Lambda=\{\alpha_1,\dots,\alpha_n\}.$
Since 
$v_1,\dots,v_n$ are nonzero constants, our task of efficiently repairing $c_i$ is reduced to finding $l$ polynomials 
$\{f_j\}_{j\in[l]}
$ of degree less than $n-k$ such that the quantity
   \begin{equation}\label{eq:m}
\sum_{t\in[n],t\neq i} {\dim}_F(\{f_j(\alpha_t)\}_{j\in[l]})
   \end{equation}
    is minimized under the 
condition that $\{f_1(\alpha_i),\dots,f_l(\alpha_i)\}$ is a basis for $E$ over $F.$

\vspace*{.1in}
Guruswami and Wootters \cite{Guruswami16} also gave a characterization for
linear repair schemes of scalar linear MDS codes based on the framework in \cite{Shanmugam14}.
We will use this characterization to prove one of our main results, namely, a lower bound on the sub-packetization, so we recall it below.  
In the next theorem $E$ is the degree-$l$ extension of the field $F$. Viewing $E$ as an $l$-dimensional vector space over $F$, we use the notation
$\dim_F(a_1,a_2,\dots,a_t)$ to refer to the dimension of the subspace spanned by the set $\{a_1,a_2,\dots,a_t\}\subset E$ over $F$.

We will need a result from \cite{Guruswami16} which we state in the form that is suited to our needs.
\begin{theorem}[\cite{Guruswami16}]\label{Thm:Guru}
Let $\cC\subseteq E^n$ be a scalar linear MDS code of length $n$. Let $F$ be a subfield of $E$ such that $[E:F]=l.$ For a given $i\in \{1,\dots,n\}$  the following statements are equivalent.

\begin{enumerate}[(1)]
\item There is a linear repair scheme of the node $c_i$ over $F$ 
such that the repair bandwidth $N(\cC,i,[n]\setminus\{i\}) \le b$.

\item There is a subset of codewords $\cP_i \subseteq \cC^{\perp}$
with size $|\cP_i|=l$ such that
$$
\dim_F ( \{x_i: x\in \cP_i\}) = l,
$$
and 
$$
b \ge \sum_{j\in[n]\setminus\{i\}} \dim_F ( \{x_j: x\in \cP_i\}).
$$
\end{enumerate}
\end{theorem}

In addition to a general linear repair scheme for scalar linear MDS codes, the authors of \cite{Guruswami16} also presented a specific repair scheme for a family of RS codes and further proved that (in some cases) the repair bandwidth of 
RS codes using this scheme is the smallest possible among all linear repair schemes and all scalar linear MDS codes with the same 
parameters. 
At the same time, the approach of \cite{Guruswami16} has some limitations. 
Namely, their repair scheme applies only for small sub-packetization $l=\log_{n/r} n$, and the optimality claim only holds for this specific sub-packetization value.  At the same time, in order to achieve the cut-set bound, $l$ needs to be exponentially large in $n$ for a fixed 
value of $r$ \cite{Goparaju14}, so the repair bandwidth of this scheme is rather far from the bound.

\section{Single-node repair: Optimal $(1,d)$ RS codes}\label{sect:single-node}

\subsection{A simple construction}\label{sect:warmup}
In this section we present a simple construction of RS codes that achieve the cut-set bound for the repair of certain nodes. 
We note that any $(n,k)$ MDS code trivially allows repair that achieves the cut-set bound for $d=k$. We say that a node in an MDS code has a {\em nontrivial optimal repair scheme} if {for a given $d>k$} it is possible to repair this node from any $d$  helper nodes with repair bandwidth achieving the cut-set bound. 
The code family presented in this section is different from standard MSR codes in the sense that although the repair bandwidth of our construction achieves the cut-set bound, the number of helper nodes depends on the node being repaired.

In the next theorem we construct a special subfamily of RS codes. Denote by $\pi(t)$ the number of primes less than or equal to $t$. Our construction enables nontrivial repair of $\pi(r)$ nodes, which
without loss of generality we take to be nodes $1,2,\dots,\pi(r)$. Let $d_i, i=1,2,\dots,\pi(r)$ be the number of helper nodes used to repair the $i$-th node. We will take $d_i=p_i+k-1$, where $p_i$ is the $i$-th smallest prime number.  The repair scheme presented below
supports repair of node $i$ by connecting to any $d_i$ helper nodes and downloading a $\frac{1}{p_i}$-th
proportion of information stored at each of these nodes. Since $p_i=d_i-k+1,$ this justifies the claim of achieving the cut-set bound for repair of a single node.

Let $m:=\pi(r)$ and let $q\ge n-m$ be a prime power. Let $E$ be the $\big(\prod_{i=1}^m p_i\big)$-th degree extension of the finite field $\mathbb{F}_q$. 

\begin{theorem} Let $n\ge k$ be two positive integers, and let $r=n-k.$ 
There exists an $(n,k)$ RS code over $E$ such that $m=\pi(r)$ of its coordinates admit nontrivial optimal repair schemes.
\end{theorem}

\begin{IEEEproof} Let $\alpha_i, i=1,\dots,m$ be an element of order $p_i$ over $\mathbb{F}_q,$ so that $\mathbb{F}_{q^{p_i}}=\mathbb{F}_q(\alpha_i),$
where $\mathbb{F}_q(\alpha_i)$ denotes the field obtained by adjoining $\alpha_i$ to $\mathbb{F}_q.$ It is clear that $E=
\mathbb{F}_q(\alpha_1,\dots,\alpha_m)$.
Define $m$ subfields $F_i$ of $E$ by setting 
     $$
     F_i=\mathbb{F}_q(\alpha_j:j\neq i),
     $$
 so that $E=F_i(\alpha_i)$ and $[E:F_i]=p_i$, $i=1,\dots,m.$
Let $\alpha_{m+1},\dots,\alpha_{n}\in \mathbb{F}_q$ be arbitrary $n-m$ distinct elements of the field, and let $\Omega=\{\alpha_1,\alpha_2,\dots,\alpha_n\}$.

Let $\cC=\text{\rm RS}_E(n,k,\Omega)$ be the RS code of dimension $k$ with evaluation points $\Omega$ and let $\cC^\bot$ be its dual code.
We claim that for $i=1,2,\dots,m,$ the $i$-th coordinate (node) of $\cC$ can be optimally repaired from any $d_i$ helper nodes, where
$$
d_i= p_i+k-1 .
$$

Let $i\in\{1,2,\dots,m\}$ and let us show how to repair the $i$th node. Choose a subset of helper nodes $\cR_i\subseteq [n]\backslash\{i\},|\cR_i|=d_i,$ and note that since $p_i\leq r,$ we have $d_i\le n-1$.
 Let $h(x)$ be the annihilator polynomial of the set $\{\alpha_j:j\in[n] \setminus(\cR_i \cup \{i\})\}$, i.e., 
\begin{equation}\label{eq:stt}
h(x)=\prod_{j\in[n] \setminus(\cR_i \cup \{i\}) } (x-\alpha_j).
\end{equation}
Since $\deg(h(x))=n-k-p_i,$ we have $\deg(x^s h(x))<r$ for all $s=0,1,\dots,p_i-1.$
As a result, for all $s=0,\dots,p_i-1$, the vector
\begin{equation}\label{eq:ed}
(v_1 \alpha_1^s h(\alpha_1),\dots,v_n \alpha_n^s h(\alpha_n))\in \cC^\bot,
\end{equation}
cf. \eqref{eq:grs}.
Let $c=(c_1,\dots,c_n)\in \cC$ be a codeword. By \eqref{eq:ed} we have
$$
\sum_{j=1}^n v_j h(\alpha_j) \alpha_j^s c_j=0, \quad s=0,\dots,p_i-1.
$$
Let $\trace_i:=\trace_{E/F_i}$ denote the trace from $E$ to $F_i.$ We have
$$
\sum_{j=1}^n\trace_i(v_j h(\alpha_j) \alpha_j^s c_j)=0, \quad s=0,\dots,p_i-1.
$$
Equivalently, we can write
\begin{equation}\label{eq:trg}
\begin{aligned}
\trace_i(v_i h(\alpha_i) \alpha_i^s c_i) & = 
- \sum_{j\neq i}\trace_i(v_j h(\alpha_j) \alpha_j^s c_j) \\
& = - \sum_{j\in \cR_i}\trace_i(v_j h(\alpha_j) \alpha_j^s c_j)  \\
& = - \sum_{j\in \cR_i}  \alpha_j^s \trace_i(v_j h(\alpha_j) c_j),
 \quad s=0,\dots,p_i-1,
\end{aligned}
\end{equation}
where the second equality follows from \eqref{eq:stt} and the third follows because $\alpha_j\in F_i$ for all $j\neq i$ and $\trace_i$ is an $F_i$-linear map.

The information used to recover the value $c_i$ (to repair the $i$th node) is comprised of the following $d_i$ elements of $F_i:$
   $$
   \trace_i(v_j h(\alpha_j) c_j), \;\;j\in\cR_i.
   $$
Let us show that these elements indeed suffice. First, by \eqref{eq:trg}, given these elements, we can calculate
the values of $\trace_{i}(v_i h(\alpha_i) \alpha_i^s c_i)$
for all $s=0,\dots,p_i-1.$ The mapping  
   $$
   \begin{array}{ll}
   E\to F_i^{p_i}\\
\gamma \mapsto \big(\trace_{i} \big(v_ih(\alpha_i)\gamma \big), 
\trace_{i} \big( v_i h(\alpha_i)\alpha_i \gamma \big) , \dots, 
\trace_{i} \big( v_i h(\alpha_i) \alpha_i^{p_i-1} \gamma \big) \big).
  \end{array}
   $$ 
is in fact a bijection, which can be realized as follows. Since the set $\{1,\alpha_i,\dots,\alpha_i^{p_i-1}\}$  forms a basis of $E$ over $F_i$
and $v_i h(\alpha_i) \neq 0,$ the set $\{v_i h(\alpha_i), v_i h(\alpha_i)\alpha_i,\dots, v_i h(\alpha_i)\alpha_i^{p_i-1}\}$
also forms a basis. 
Let $\{ \theta_0, \theta_1, \dots, \theta_{p_i-1} \}$ be the dual basis of $\{v_i h(\alpha_i), v_i h(\alpha_i)\alpha_i,\dots, v_i h(\alpha_i)\alpha_i^{p_i-1}\},$ i.e., 
$$
\trace_i(v_i h(\alpha_i)\alpha_i^s  \theta_j) = \left\{
\begin{array}{cc}
0, & \text{~if~} s \neq j \\
1, & \text{~if~} s = j 
\end{array} \right. 
\text{for all~} s,j\in\{0,1,\dots,p_i-1\}.
$$
According to \eqref{eq:expansion}, the value $c_i$ can now be found as follows:
$$
c_i= \sum_{s=0}^{p_i-1} \trace_i(v_i h(\alpha_i)\alpha_i^s c_i) \theta_s.
$$

The presented arguments constitute a linear repair scheme of the node $c_i, i=1,\dots m$ over $F_i.$ 
The information downloaded from each of the helper nodes consists of one element
of $F_i,$ or, in other words, the $(1/p_i)$th proportion of the contents of each node.
This shows that node $i$ admits nontrivial optimal repair. 
The proof is thereby complete.
\end{IEEEproof}

\begin{example} Take $q=5,$ $k=3,r=5$. We have $\pi(r)=3$ and $p_1=2, p_2=3, p_3=5.$ 
Let us construct an $(8,3)$ RS code over the field $E=\mathbb{F}_{5^{30}}$, where the first $3$ nodes admit  nontrivial optimal repair schemes. Let  $\alpha$ be a primitive element of $E$. Choose the set $\Omega=\{\alpha_1,\dots,\alpha_8\}$ as follows:
   $$\alpha_1=\alpha^{\frac{5^{30}-1}{5^2-1}},\alpha_2=\alpha^{\frac{5^{30}-1}{5^3-1}},\alpha_3=\alpha^{\frac{5^{30}-1}{5^5-1}},\alpha_4=0,\alpha_5=1,\alpha_6=2,\alpha_7=3,\alpha_8=4.
   $$
  The number of helper nodes for the first $3$ nodes is $(d_1,d_2,d_3)=(4,5,7)$.
It is easy to verify that for any subset  $A\subseteq \{1,2,3\}$
$$ \mathbb{F}_5(\alpha_i:i\in A)=\mathbb{F}_{m_{\!_A}}, \text{ where } m_{_A}=5^{(\prod_{i\in A}p_i)}.$$

The code $\cC$ constructed in the above proof is given by $\cC=\text{\rm RS}_E(8, 3,\Omega).$ Let us address the task
of repairing $c_3$ from all the remaining $7$ helper nodes with repair bandwidth achieving the cut-set bound.
Let $\cC^\bot=\text{\rm GRS}_E(8, 5,\Omega,v)$, where 
$v=(v_1,\dots,v_8)\in (E^\ast)^8$.
We download the value $\trace_{E/\mathbb{F}_{5^{6}}}(v_jc_j)$ from each helper node $c_j,j\neq 3$.
Since $[E:\mathbb{F}_{5^{6}}]=p_3,$ this amounts to downloading exactly a $1/p_3=(1/5)$-th fraction
of the information stored at each helper node, which achieves the cut-set bound.
The value of $c_3$ can be found from the downloaded information using the following $5$ equations: 
    $$
  \trace_{E/\mathbb{F}_{5^{6}}}(\alpha_3^s v_3c_3) =
 - \sum_{j\neq 3}\trace_{E/\mathbb{F}_{5^{6}}}(\alpha_j^s v_jc_j)=
 - \sum_{j\neq 3}\alpha_j^s\trace_{E/\mathbb{F}_{5^{6}}}(v_jc_j)
, \quad  s=0,\dots,4.
$$ 
Indeed, the downloaded symbols suffice to recover the vector $(\trace_{E/\mathbb{F}_{5^{6}}}(\alpha_3^s v_3c_3),{s=0,\dots,4})$, and
therefore also suffice to repair the symbol $c_3$. 
\end{example}

\subsection{The basic field tower}\label{Sect:tower}
The code constructions and repair schemes that we define are based on a tower of field extensions shown in Figure 1. In this section we
give a general definition of the tower that will be used in Sections \ref{Sect:cons} ,\ref{sect:warmup-multiple}, and \ref{Sect:uee} below.

%

Let $\mathbb{F}_p$ be a finite field (for simplicity we can take $p=2$) and let $s$ be a natural number
whose value will be specified later.

Let $p_1,\dots,p_n$ be $n$ distinct primes such that
\begin{equation}\label{eq:pmsm}
p_i\equiv 1 \;\text{mod}\, s \;\;\text{~for all~} i=1,2,\dots,n.
\end{equation}
According to Dirichlet's theorem, there are infinitely many such primes. For $i=1,\dots,n$, let $\alpha_i$ be an element of 
degree $p_i$ over $\mathbb{F}_p$, i.e., $[\mathbb{F}_p(\alpha_i):\mathbb{F}_p]=p_i$, and define the fields $F_i=\ff_p(\alpha_j,j\in [n]\backslash\{i\}), i=1,2,\dots,n.$ For a given $i\in[n]$, the field $F_i$ contains all the $\alpha_j$ except $\alpha_i.$
Adjoining $\alpha_i$ to $F_i$, we obtain the field
\begin{equation}\label{eq:defFm}
\mathbb{F}:=\mathbb{F}_p(\alpha_1,\dots,\alpha_n).
\end{equation}
Note that for any subset of indices $A\subseteq [n]$, the field $\mathbb{F}_p(\{\alpha_i:i\in A\})$ is an extension of $\mathbb{F}_p$
of degree $\prod_{i\in A}p_i,$ and in particular, $\mathbb{F}$ has degree $\prod_{i=1}^np_i$ over $\mathbb{F}_p$. 
For $i=1,\dots,n$ 

Finally, let $\mathbb{K}$ be an algebraic extension of $\mathbb{F}$ of degree $s$ and let $\beta\in \mathbb{K}$ 
be such that 
\begin{equation}\label{eq:bbk}
\mathbb{K}=\mathbb{F}(\beta)
\end{equation}
($\beta$ always exists by the primitive element theorem).

%
%

\vspace*{.2in}
\noindent\fbox{\parbox{.95\linewidth}
{\begin{center}
\begin{tikzpicture}[scale=1.3]
\draw node at (9,-1.5) [text width=0.3in, align=center](A) {$\ff_p$};
\draw node at (6,0) [text width=0.3in, align=center](B) {$F_1$};
\draw node at (8,0) [text width=0.3in, align=center](C) {$F_2$};
\draw node at (10,0) [text width=0.3in, align=center](D) {$\dots$};
\draw node at (12,0) [text width=0.3in, align=center](E) {$F_n$};
\draw[->, >=stealth,line width=.2mm] (A) -- (B) node[draw=none,fill=none,font=\scriptsize,midway,below] {$\tilde\alpha_1$};
\draw[->, >=stealth,line width=.2mm] (A) -- (C) node[draw=none,fill=none,font=\scriptsize,near start,above] {$\;\;\tilde\alpha_2$};
\draw[->, >=stealth,line width=.2mm] (A) -- (E) node[draw=none,fill=none,font=\scriptsize,midway,below] {$\tilde\alpha_n$};
\draw node at (9,1.5) [text width=0.3in, align=center](F) {$\ff$};
\draw[->, >=stealth,line width=.2mm] (B) -- (F) node[draw=none,fill=none,font=\scriptsize,midway,below] {$\hspace*{.2in}(\alpha_1,p_1)$};
\draw[->, >=stealth,line width=.2mm] (C) -- (F) node[draw=none,fill=none,font=\scriptsize,midway,below] {$\hspace*{.35in}(\alpha_2,p_2)$};
\draw[->, >=stealth,line width=.2mm] (E) -- (F) node[draw=none,fill=none,font=\scriptsize,midway,below] {\hspace*{-.2in}$(\alpha_n,p_n)$};
\draw node at (9,2.8) [text width=0.3in, align=center](G) {$\mathbb{K}$};
\draw[->, >=stealth,line width=.2mm] (F) -- (G) node[draw=none,fill=none,font=\scriptsize,near end,below] {$\hspace*{.35in}(\beta,s)$};
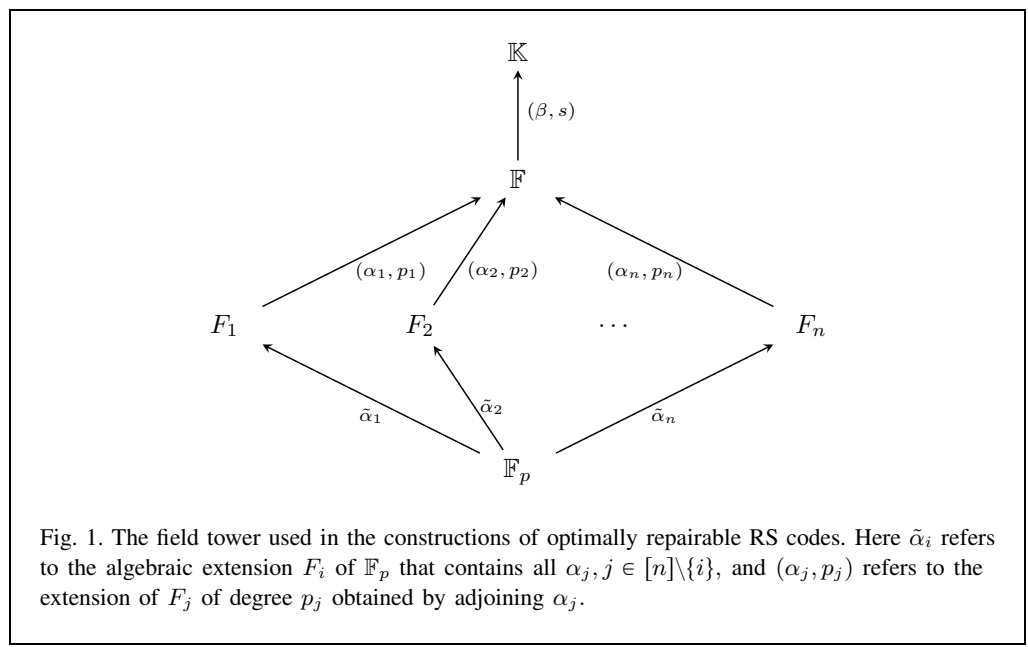
\draw node at (9,-2.5) [text width=5in, align=left] {\small Fig. 1. The field tower used in the constructions of optimally repairable
RS codes. Here
$\tilde \alpha_i$ refers to the algebraic extension $F_i$ of $\ff_p$ that contains all $\alpha_j, j\in[n]\backslash\{i\},$ and $(\alpha_j,p_j)$ refers to the extension of $F_j$ of degree $p_j$ obtained by adjoining $\alpha_j.$};
\end{tikzpicture}
\end{center}
}}

\vspace*{.2in}

\subsection{A family of RS codes achieving the cut-set bound}\label{Sect:cons}

In this section we develop the ideas discussed above and construct  RS codes achieving the cut-set bound with
nontrivial optimal repair of all nodes. More precisely, given any positive integers $k< d \leq n-1$, we explicitly construct an $(n,k)$ RS code $\cC$ achieving the cut-set bound for the repair of {\em any} single node from {\em any} $d$ helper nodes. In other words, $\cC$ is an $(n,k)$ MSR code with repair degree $d$.

The codes are constructed using the set of evaluation points $\alpha_1,\alpha_2\dots,\alpha_n$ defined in Sec.~\ref{Sect:tower}. Throughout this section we set $s=d-k+1$ (cf.~\eqref{eq:pmsm}, \eqref{eq:bbk}).
Before stating the main result, we note that the condition required of $\alpha_1,\alpha_2\dots,\alpha_n$
is of the form $\alpha_i\not \in \mathbb{F}_q(\alpha_j,j\ne i), i=1,\dots,n$. The most efficient way to accomplish this in terms of the value of sub-packetization $l$ is to take the extension degrees to be the smallest (distinct) primes, and this is the underlying idea behind the sub-family of the RS codes that we consider. The new element in the construction compared with Sec.~\ref{sect:warmup} above,
that enables the repair of all nodes, is the introduction of the extension field $\mathbb{K}.$

The following theorem is the main result of this section.
\begin{theorem} \label{thm1} 
Let $k,n,d$ be any positive integers such that $k< d < n.$ Let $\Omega=\{\alpha_1,\dots,\alpha_n\}$, where $\alpha_i,i=1,\dots,n$ is an element of degree $p_i$
over $\mathbb{F}_p$ and $p_i$ is the $i$th smallest prime that satisfies \eqref{eq:pmsm}.
 The code $\cC:=\text{\rm RS}_{\mathbb{K}}(n,k,\Omega)$ achieves the cut-set bound for the repair of any single node from any $d$ helper nodes.
In other words, $\cC$ is an $(n,k)$ MSR code with repair degree $d$. 
 \end{theorem}
\begin{IEEEproof}
Our repair scheme of the $i$-th node is performed over the field $F_i$. More specifically, for every $i\in[n]$, we explicitly construct a vector space $S_i$ over the field $F_i$ such that
\begin{equation}\label{eq:Si}
\dim_{F_i} S_i = p_i, \quad
  S_i + S_i\alpha_i+\dots + S_i\alpha_i^{s-1}=\mathbb{K},
\end{equation}
where $S_i \alpha := \{\gamma \alpha: \gamma \in S_i\}$,
and the operation $+$ is the Minkowski sum of sets, $T_1 + T_2 := \{\gamma_1+\gamma_2:\gamma_1\in T_1, \gamma_2\in T_2 \}.$ Note that the sum in \eqref{eq:Si} is in fact a direct sum since the dimension of each summand is $p_i$, and $[\mathbb{K}:F_i]=sp_i$.
We will describe a construction of $S_i$ and prove that $S_i$ satisfies \eqref{eq:Si} in Lemma~\ref{lem1} later in this section. For now let us assume that we have such vector spaces $S_i,i=1,2,\dots,n$ and continue the proof of the theorem.

Suppose that we want to repair the $i$-th node from a subset $\cR\subseteq [n]\backslash\{i\}$ of 
$|\cR|=d$ helper nodes. 
Let $h(x)$ be the annihilator polynomial of the set $\{\alpha_j: j\in[n]\setminus (\cR\cup\{i\}) \}$, i.e., 
\begin{equation}\label{eq:defh}
h(x)=\prod_{j\in[n]\setminus (\cR\cup\{i\})}(x-\alpha_j).
\end{equation}
By \eqref{eq:grs} the dual code of $\cC$ is $\cC^\bot=\text{\rm GRS}_{\mathbb{K}}(n,n-k,\Omega,v)$ where the coefficients  
$v=(v_1,\dots,v_n) \in (\mathbb{K}^*)^n$ are nonzero.
Clearly, $\deg(x^th(x))\leq s-1+n-(d+1)<n-k$ for all $t=0,1,\dots,s-1,$ so for any such $t$ we have
\begin{equation}\label{eq:dual}
(v_1\alpha_1^t h(\alpha_1),\dots,v_n \alpha_n^t h(\alpha_n))\in\cC^\bot.
\end{equation}
These $s$ dual codewords will  be used to recover the $i$-th coordinate. 
Let $c=(c_1,\dots,c_n)\in \cC$ be a codeword, and let us construct a repair scheme for the coordinate (node) $c_i$ using
the values $\{c_j:j\in \cR\}$. Rewrite \eqref{eq:dual} as follows:
\begin{equation}\label{eq:inter}
\sum_{j=1}^n  v_j\alpha_j^t h(\alpha_j) c_j =0 \text{~for all~}  t=0,\dots,s-1.
\end{equation}
Let $e_1,\dots,e_{p_i}$ be an arbitrary basis of the subspace $S_i$ over the field $F_i$. From \eqref{eq:inter} we obtain the
following system of $sp_i$ equations:
$$
\sum_{j=1}^n  e_m v_j\alpha_j^t h(\alpha_j) c_j =0,\;\;   t=0,\dots,s-1;
m=1,\dots,p_i.
$$
Let $\trace_i:=\trace_{\mathbb{K}/F_i}$ be the trace map to the subfield $F_i$. From the last set of equations we have
\begin{equation}\label{eq:2}
\sum_{j=1}^n\trace_i(e_mv_j\alpha_j^th(\alpha_j)c_j)=0 \text{~for all~} t=0,\dots,s-1 \text{~and all~} m=1,\dots,p_i,
\end{equation}
Arguing as in \eqref{eq:trg}, let us write \eqref{eq:2} in the following form:
\begin{equation}\label{eq:rcv}
\begin{aligned}
\trace_i(e_m \alpha_i^t v_i h(\alpha_i)c_i) & = - \sum_{j\neq i} \trace_i(e_mv_j\alpha_j^th(\alpha_j)c_j) \\
& = - \sum_{j\in \cR} \trace_i(e_m v_j \alpha_j^t h(\alpha_j)c_j) \\
& = - \sum_{j\in \cR} \alpha_j^t h(\alpha_j)\trace_i(e_m v_j  c_j)
\end{aligned}
\end{equation}
for all $t=0,\dots,s-1$ and all $m=1,\dots,p_i,$
where the second equality follows from \eqref{eq:defh} and the third follows from the fact that the trace mapping $\trace_i$ is $F_i$-linear, and that $\alpha_j\in F_i$ for all $j \neq i$.

As before, to recover $c_i$, we download the following $p_i$ symbols of $F_i$ from each helper node  $c_j, j\in \cR$:
\begin{equation}\label{eq:dl}
\trace_i(e_m v_j  c_j) \text{~for~} m=1,\dots,p_i.
\end{equation}
These field elements suffice to recover the node $c_i$. Indeed, according to \eqref{eq:rcv}, we can calculate the values of 
$\trace_{i}(e_m \alpha_i^t v_i h(\alpha_i)c_i)$ for all $t=0,\dots,s-1$ and all $m=1,\dots,p_i$ from the set of elements in \eqref{eq:dl}.
By definition, $e_1,\dots,e_{p_i}$ is a basis of the subspace $S_i$ over the field $F_i$.
According to \eqref{eq:Si}, {
 $\mathbb{K} = S_i + S_i\alpha_i+\dots + S_i\alpha_i^{s-1}$.  }
Therefore, the set $\{e_m\alpha_i^t:  t=0,\dots,s-1 ;\, m=1,\dots,p_i\}$ forms a basis of  $\mathbb{K}$ over  $F_i$
and so does the set  $\{e_m\alpha_i^t v_i h(\alpha_i):  t=0,\dots,s-1 ;\, m=1,\dots,p_i\}$ (recall that $v_i h(\alpha_i)\neq 0$).
Hence the mapping 
 $$
   \begin{array}{ll}
   \mathbb{K}\to F_i^{s p_i}\\
\gamma \mapsto (\trace_{i}(e_m \alpha_i^t v_i h(\alpha_i) \gamma), m=1,\dots,p_i; t=0,\dots,s-1).
  \end{array}
   $$ 
 is a bijection. 
This means that $c_i$ is uniquely determined by the set of values $\{\trace_{i}(e_m \alpha_i^t v_i h(\alpha_i) c_i),m=1,\dots,p_i; t=0,\dots,s-1\}$, validating our repair scheme.

It is also clear that the construction meets the cut-set bound. Indeed, $c_j\in\mathbb{K}$ for all $j$
 and $[\mathbb{K}:F_i]=sp_i,$ so
the amount of information required from each helper node \eqref{eq:dl} is exactly $(1/s)$th fraction of its contents.

This completes the proof of Theorem \ref{thm1}.
\end{IEEEproof}

In the proof above we assumed the existence of the vector space $S_i$ that satisfies \eqref{eq:Si} for all $i\in[n]$.
In the next lemma we construct such a space and establish its properties. 


For a vector space $V$ over a field $F$ and a set of vectors $A=(a_1,\dots,a_l)\subset V$, let $\spun_F(A)=\{\sum_{i=1}^l \gamma_ia_i, \gamma_i\in F\}$ be the span of $A$
 over $F$.

\begin{lemma}\label{lem1} 
Let $\beta$ be a generating element of $\mathbb K$ over $\mathbb{F}=\mathbb{F}_p(\alpha_1,\dots,\alpha_n).$
Given $i\in [n]$, define the following vector spaces over $F_i$:
\begin{gather*}
S_i^{(1)} =\spun_{F_i} \big(\beta^u \alpha_i^{u+qs}, u=0,1,\dots,s-1; q=0,1,\dots,\textstyle{\frac{p_i-1}{s}}-1 \big) \\
S_i^{(2)}  =\spun_{F_i} \Big(\sum_{t=0}^{s - 1}\beta^t \alpha_i^{p_i-1} \Big) \\
S_i   =S_i^{(1)} + S_i^{(2)}.
\end{gather*}
Then 
$$
\dim_{F_i} S_i = p_i, \quad
S_i + S_i\alpha_i+\dots + S_i\alpha_i^{s-1}=\mathbb{K}.
$$
\end{lemma}

\begin{IEEEproof} Let {$K:=S_i + S_i\alpha_i+\dots + S_i\alpha_i^{s-1}$.} If $K=\mathbb{K},$ then $\dim_{F_i} S_i = p_i$
easily follows. Indeed, by definition $\dim_{F_i} S_i \le p_i$. On the other hand, $[\mathbb{K}:F_i]=s p_i$ and
{$K=\mathbb{K}$} together imply that $\dim_{F_i} S_i \ge p_i$. 

Let us prove that $K=\mathbb{K}.$ Clearly $K$ is a vector space over $F_i$, and $K\subseteq \mathbb{K}$. 
Let us show the reverse inclusion, namely that $\mathbb{K}\subseteq K$. To prove this, recall that 
$\mathbb{K}$ is a vector space of dimension $s$ over $\mathbb{F}$ (see \eqref{eq:bbk}), and the set $1,\beta,\dots,\beta^{s-1}$
forms a basis, i.e.,  $\mathbb{K}=\oplus_{u=0}^{s-1}\beta^u\mathbb{F}$.
Thus, the lemma will be proved if we show that 
$\beta^u\mathbb{F} \subseteq K$ for all $u=0,1,\dots,s-1.$ 
To prove this inclusion we will use induction on $u$.

For the induction base, let $u=0$. 
In this case, we have $\alpha_i^{qs} \in S_i^{(1)}$ for all $0\leq q<\frac{p_i-1}{s}$.
Therefore $\alpha_i^{qs+j} \in S_i^{(1)} \alpha_i^j$ for all $0\leq q<\frac{p_i-1}{s}$.
As a result, $\alpha_i^{qs+j} \in K$ for all $0\le q<\frac{p_i-1}{s}$ and all $0 \le j \le s-1$.
In other words, 
\begin{equation}\label{eq:u1}
\alpha_i^t \in K ,\; t=0,1,\dots, p_i-2.
\end{equation}

Next we show that also $\alpha_i^{p_i-1} \in K$.
For every $t=1,\dots,s-1$ we have $0\le \lfloor \frac{p_i-1-t}{s} \rfloor <\frac{p_i-1}{s}$.
As a result,
$$
\beta^t \alpha_i^{t+ \lfloor \frac{p_i-1-t}{s} \rfloor s} \in S_i^{(1)}, \;
t=1,\dots,s-1.
$$
We obtain, for each $t=1,\dots,s-1,$
$$
\beta^t \alpha_i^{p_i-1} = 
\beta^t \alpha_i^{t+ \lfloor \frac{p_i-1-t}{s} \rfloor s} 
\alpha_i^{p_i-1-t - \lfloor \frac{p_i-1-t}{s} \rfloor s}
 \in S_i  \alpha_i^{p_i-1-t - \lfloor \frac{p_i-1-t}{s} \rfloor s} \subseteq K.
$$
At the same time,
$$
\sum_{t=0}^{s - 1}\beta^t \alpha_i^{p_i-1} \in S_i^{(2)} \subseteq K.
$$
 The last two statements together imply that
$$
\alpha_i^{p_i-1} = \sum_{t=0}^{s - 1}\beta^t \alpha_i^{p_i-1}
- \sum_{t=1}^{s - 1}\beta^t \alpha_i^{p_i-1}  \in K.
$$
Combining this with \eqref{eq:u1}, we conclude that
$\alpha_i^t \in K$  for all $t=0,1,\dots, p_i-1$.
Recall that $1,\alpha_i,\dots,\alpha_i^{p_i-1}$  is a basis of $\mathbb{F}$ over $F_i$, and that
$K$ is a vector space over $F_i$, so $\mathbb{F} \subseteq K$.
This establishes the induction base.

Now let us fix $u\ge 1$ and let us assume that $\beta^{u'}\mathbb{F} \subseteq K$ for all $u'<u.$ To prove the induction step, we need to show that $\beta^{u}\mathbb{F}\subseteq K$.
Mimicking the argument that led to \eqref{eq:u1}, we can easily show that
\begin{equation}\label{eq:ubeta}
\beta^u\alpha_i^{u+t} \in K,\; t=0,1,\dots, p_i-2.
\end{equation}
Let us show that \eqref{eq:ubeta} is also true for $t=p_i-1,$ i.e., that $\beta^u\alpha_i^{u+p_i-1} \in K$.
For every $1 \le t \le s-1-u$, we have $0\le \lfloor \frac{p_i-1-t}{s} \rfloor <\frac{p_i-1}{s}$.
As a result,
$$
\beta^{u+t} \alpha_i^{u+t+ \lfloor \frac{p_i-1-t}{s} \rfloor s} \in S_i^{(1)},\; t=1,\dots, s-1-u.
$$
Therefore, for all such $t$
\begin{equation}\label{eq:c1}
\beta^{u+t} \alpha_i^{u+p_i-1} = 
\beta^{u+t} \alpha_i^{u+t+ \lfloor \frac{p_i-1-t}{s} \rfloor s} 
\alpha_i^{p_i-1-t - \lfloor \frac{p_i-1-t}{s} \rfloor s} 
  \in S_i  \alpha_i^{p_i-1-t - \lfloor \frac{p_i-1-t}{s} \rfloor s} \subseteq K
\end{equation}
By the induction hypothesis, $\beta^{u'} \mathbb{F} \subseteq K$ for all $u'=0,1,\dots,u-1.$ As a result,
\begin{equation}\label{eq:c2}
\beta^{u'} \alpha_i^{u+p_i-1} \in K,\; u'=0,1,\dots,u-1.
\end{equation}
At the same time,
\begin{equation}\label{eq:c3}
\sum_{t=0}^{s - 1}\beta^t \alpha_i^{u+p_i-1}
= \Big( \sum_{t=0}^{s - 1}\beta^t \alpha_i^{p_i-1} \Big) \alpha_i^u
\in S_i^{(2)} \alpha_i^u \subseteq K.
\end{equation}
Combining \eqref{eq:c1}, \eqref{eq:c2} and \eqref{eq:c3}, we obtain
$$
\beta^u\alpha_i^{u+p_i-1}
= \sum_{t=0}^{s - 1}\beta^t \alpha_i^{u+p_i-1}
- \sum_{u'=0}^{u-1} \beta^{u'} \alpha_i^{u+p_i-1}
- \sum_{t=1}^{s-1-u} \beta^{u+t} \alpha_i^{u+p_i-1}
\in K.
$$
Now on account of \eqref{eq:ubeta} we can conclude that
$\beta^u\alpha_i^{u+t} \in K$  for all $ t =0,1,\dots, p_i-1.$
Therefore, $\beta^u\mathbb{F} \subseteq K$.
This establishes the induction step and completes the proof of the lemma.
\end{IEEEproof}

The value of sub-packetization of the constructed codes is given in the following obvious proposition.

\begin{proposition}
The sub-packetization of our construction is 
$l=[\mathbb{K}:\mathbb{F}_p]=s\prod_{i=1}^n p_i$, where the $p_i$'s are the smallest $n$ distinct primes satisfying \eqref{eq:pmsm}. 
\end{proposition}
The proof follows immediately from the fact that the repair of the $i$-th coordinate is performed over the field $F_i$, so the repair field of our construction is $\cap_{i=1}^n F_i = \mathbb{F}_p$. To estimate the asymptotics of $l$ for $n\to\infty,$ 
recall that our discussion of Dirichlet's prime number theorem in Sec.~\ref{sect:results} above implies that, for fixed $s$, $l= e^{(1+o(1)) n\log n}$. This proves the upper bound in \eqref{eq:main}.

\section{A lower bound on the sub-packetization of scalar linear MSR codes}\label{Sect:lb}
In this section we prove a lower bound on the sub-packetization value $l$ of $(n,k)$ scalar linear MSR codes, which implies that 
$l\ge e^{(1+o(1))k \log k}$. In contrast, for MSR array codes, a much smaller sub-packetization value $l=r^{\lceil n/(r+1) \rceil}$ is achievable \cite{Wang16}. This shows that limiting oneself to scalar linear  codes necessarily leads to a much larger
sub-packetization, and constructing such codes in real storage systems is even less feasible than their array code counterparts. 
The main result of this section is the following theorem:
\begin{theorem}\label{thm2} Let $F=\mathbb{F}_{q}$ and $E=\mathbb{F}_{q^l}$ for a prime power $q$.
{Let $d$ be an integer between $k+1$ and $n-1$.}
Let ${\cC} \subseteq E^n$ be an $(n,k)$ scalar linear MDS code with a linear repair scheme over $F.$ Suppose that the repair bandwidth of the scheme
achieves the cut-set bound with equality for the repair of any single node from any $d$ helper nodes. Then the sub-packetization $l$ is at least  
    $$ l \geq \prod_{i=1}^{k-1}p_i
    $$ 
    where $p_i$ is the $i$-th smallest prime.  
\end{theorem} 

As discussed above in Sec. \ref{sect:results}, this theorem implies the asymptotic lower bound  $l\ge e^{(1+o(1))k \log k}.$

\vspace*{.1in} In the proof of Theorem \ref{thm2}, we will need the following auxiliary lemmas.
\begin{lemma}\label{lem:subfield} {\rm (Subfield criterion \cite[Theorem 2.6]{Lidl94})} Each subfield  of the field $\ff_{p^n}$ is of order $p^m,$ where $m|n.$
For every positive divisor $m$ of $n$ there exists a unique subfield of $\ff_{p^n}$ that contains $p^m$ elements.
\end{lemma}

\begin{lemma}\label{lem2} Let $E$ be an extension field of  $\ff_q$ and let $\alpha_1,\dots,\alpha_n\in E.$ Then
$$[\ff_q(\alpha_1,\dots,\alpha_n):\ff_q]=\lcm(d_1,\dots,d_n),$$
where $d_i=[\ff_q(\alpha_i):\ff_q].$
\end{lemma}
Proof: Obvious.

\begin{lemma}\label{lem:ez} 
Let $a_1,a_2,\dots,a_n\in F^m$ and $b_1,b_2,\dots,b_n\in F^m$
be two sets of vectors over a field $F$, and let $A$ and $B$ denote their spans over $F$. 
Let $c_i=a_i+b_i, i=1,\dots,n$ then
\begin{equation}\label{eq:plus}
\dim_F(c_1,\dots,c_n) \le \dim A + \dim B.
\end{equation}
\end{lemma}
The lemma follows immediately from the fact that, for any two subspaces $A$ and $B$ of a linear space,
\begin{equation}\label{eq:sumcap}
\dim(A+B)+\dim(A\cap B)=\dim A+\dim B.
\end{equation}
%

In the next lemma $\cS_F(\cdot)$ refers to the row space of the matrix argument over the field $F$.

\begin{lemma}\label{lem:use}
Let $E$ be an extension of a finite field $F$ of degree $l$.
Let $A=(a_{i,j})$ be an $m\times n$ matrix over $E$. Then
  \begin{equation}\label{eq:dz}
\dim(\cS_F(A))
\le \sum_{j=1}^n \dim_F(a_{1,j},a_{2,j},\dots,a_{m,j}).
  \end{equation}
Moreover, if \eqref{eq:dz} holds with equality, then for every $\cJ \subseteq [n],$
  \begin{equation}\label{eq:dzcon}
\dim(\cS_F(A_{\cJ}))
= \sum_{j\in\cJ} \dim_F(a_{1,j},a_{2,j},\dots,a_{m,j})
  \end{equation}
where $A_{\cJ}$ is the restriction of $A$ to the columns with indices in $\cJ$.
\end{lemma}
\begin{IEEEproof} Inequality \eqref{eq:dz} is an immediate consequence of Lemma \ref{lem:ez}. Indeed, suppose that $n=2$ and
view the $i$th row of $A$ as the sum of two $2$-dimensional vectors over $E$, namely $(a_{i,1}|0)$ and $(0|a_{i,2}),i=1,\dots,m;$
then \eqref{eq:dz} is the same as \eqref{eq:plus}. The extension to $n>2$ follows by straightforward induction. 

Now let us prove the second part of the claim. Suppose that 
   $$
   \dim(\cS_F(A))= \sum_{j=1}^n \dim_F(a_{1,j},a_{2,j},\dots,a_{m,j}).
   $$
Then for every $\cJ \subseteq [n],$
\begin{align*}
& \sum_{j\in\cJ} \dim_F(a_{1,j},a_{2,j},\dots,a_{m,j})
+ \sum_{j\in\cJ^c} \dim_F(a_{1,j},a_{2,j},\dots,a_{m,j}) \\ 
= & \dim(\cS_F(A))
\le \dim(\cS_F(A_{\cJ})) + \dim(\cS_F(A_{\cJ^c})).
\end{align*}
But according to \eqref{eq:dz},
\begin{align*}
\dim(\cS_F(A_{\cJ})) \le \sum_{j\in\cJ} \dim_F(a_{1,j},a_{2,j},\dots,a_{m,j}), \\
\dim(\cS_F(A_{\cJ^c})) \le \sum_{j\in\cJ^c} \dim_F(a_{1,j},a_{2,j},\dots,a_{m,j}).
\end{align*}
Therefore
$$
\dim(\cS_F(A_{\cJ})) = \sum_{j\in\cJ} \dim_F(a_{1,j},a_{2,j},\dots,a_{m,j}).
$$
This completes the proof of the lemma.
\end{IEEEproof}

Now we are ready to prove Theorem \ref{thm2}.
{}

\vspace*{.1in}{\em Proof of Theorem \ref{thm2}:}
Let $\cC$ be an $(n,k)$ MSR code with repair degree $d.$ By puncturing the code $\cC$ to any $d+1$ coordinates, we obtain a $(d+1,k)$ MSR code with repair degree $d$.
Therefore without loss of generality below we assume that $d=n-1$.

Let $H=[M | I_r]$ be the parity-check matrix of the code ${\cC}$ over $E$, written in systematic form, where $M$ is an $r\times k$ matrix and $I_r$ is the $r\times r$ identity matrix.  Let $h_{ij}$ be the entry of $H$ in position $(i,j).$ Since ${\cC}$ is an MDS code, every square submatrix of $M$ is invertible. In particular, every entry of $M$ is nonzero, so without loss of generality we may assume that $h_{1,j}=1, j=1,2,\dots, k.$
Since $d\ge k+1$, we also have $n\ge k+2$, and therefore $H$ contains at least two rows.

The theorem will follow from the following claim. 

\begin{claim}\label{claim1}
 For $j=1,\dots,k-1$ define $\alpha_j := \frac{h_{2,j}}{h_{2,k}}$.  Then for every $j=1,\dots,k-1$, 
  \begin{equation}\label{eq:claim}
\alpha_j\notin \mathbb{F}_q \big( \big\{ \alpha_i:i\in \{1,2,\dots,k-1\}\setminus\{j\} \big\} \big).
  \end{equation}
 In other words,  $\alpha_j$ is not generated by the remaining  $\alpha_i$'s over $\mathbb{F}_q$.
\end{claim}

We first show that this claim indeed implies the theorem.  
Let $d_i=[\mathbb{F}_q(\alpha_i):\mathbb{F}_q]$ be the degree of the field extension generated by 
$\alpha_i$. We prove by contradiction that for all $j=1,2,\dots,k-1$, $d_j$ does not divide 
$\lcm(d_i:i\in \{1,2,\dots,k-1\} \setminus \{j \})$. Suppose the contrary, i.e., that there is a $j$ such that $d_j|\lcm(d_i:i\in \{1,2,\dots,k-1\} \setminus \{j\}).$ According to Lemma~\ref{lem2},
$$
[\mathbb{F}_q \big( \big\{ \alpha_i:i\in \{1,2,\dots,k-1\}\setminus\{j\} \big\} \big) : \mathbb{F}_q]
= \lcm(d_i:i\in \{1,2,\dots,k-1\} \setminus \{j \}).
$$
Then by Lemma \ref{lem:subfield}, there is a subfield 
    \begin{equation}\label{eq:fj}
F_j \subseteq \mathbb{F}_q \big( \big\{ \alpha_i:i\in \{1,2,\dots,k-1\}\setminus\{j\} \big\} \big) 
    \end{equation}
such that $[F_j :\mathbb{F}_q]=d_j$. Notice that $E=\mathbb{F}_{q^l}$ contains all $\alpha_u, u=1,2,\dots,k-1$.
So both $F_j$ and $\mathbb{F}_q(\alpha_j)$ are subfields of $E$, and they have the same order $q^{d_j}$.
Consequently, $\mathbb{F}_q(\alpha_j) = F_j$. Then from \eqref{eq:fj} we conclude that 
$\alpha_j\in\mathbb{F}_q \big( \big\{ \alpha_i:i\in \{1,2,\dots,k-1\}\setminus\{j\} \big\} \big),$ which contradicts 
\eqref{eq:claim}. Thus, our assumption is wrong, and $d_j$ does not divide $\lcm(d_i:i\in \{1,2,\dots,k-1\} \setminus \{j \})$. 
As an immediate corollary,
$$
l= [E:\mathbb{F}_q] \ge
[\mathbb{F}_q(\{\alpha_i:i=1,\dots,k-1\}):\mathbb{F}_q]=\lcm(d_1,\dots,d_{k-1})\geq \prod_{i=1}^{k-1}p_i.
$$ 
Thus we have shown that Claim~\ref{claim1} indeed implies the theorem.  A proof of the claim is given in Appendix~\ref{ap:claim}.

\section{Optimal repair of two erasures}\label{sect:warmup-multiple}
In this section we present an explicit construction of RS codes that achieve the cut-set bound \eqref{eq:cs1} for the repair of any two failed nodes.

\subsection{Code construction}
Our construction is based on the field tower defined in Sec.~\ref{Sect:tower} above. In this section we take $s=s_1s_2,$ where
\begin{equation}\label{eq:s12}
s_1=d+1-k, \quad s_2=d+2-k.
\end{equation}

Let us fix the values of the code length $n$ and dimension $k.$ Let $d, k\le d\le n-2$ be the number of helper nodes used for recovery.  In the case of $h=2$ the cut-set bound \eqref{eq:cutset} has the form $\beta(2,d)=\frac{2dl}{d+2-k}.$
Our goal will be accomplished if we construct codes and a repair procedure that relies on downloading a ${2}/(d+2-k)$ fraction of the node contents from each of the helper nodes.

The codes that we construct have length $n$ and use $\{\alpha_1,\dots,\alpha_n\}$ as the set of evaluation points.
Our results are summarized in the following theorem.
\begin{theorem} \label{thm:2err} 
Let $k,n,d$ be any positive integers such that $k< d < n.$ Let $\Omega=\{\alpha_1,\dots,\alpha_n\}$, where $\alpha_i,i=1,\dots,n$ is an element of degree $p_i$ over $\mathbb{F}_p$ and $p_i$ is the $i$th smallest prime that satisfies \eqref{eq:pmsm}.
 Then the code $\cC:=\text{\rm RS}_{\mathbb{K}}(n,k,\Omega)$ 
 has the $(2,d)$-optimal repair property. 

The sub-packetization value of the code $\cC$ equals
   \begin{equation}\label{eq:2sp}
   l=[\mathbb{K}:\mathbb{F}_p]=s \prod_{i=1}^n p_i.
   \end{equation}
   For fixed $r$ and growing $n$ we have $l=e^{(1+o(1))n \log n}.$
 \end{theorem}
\begin{IEEEproof}
We write a codeword of $\cC$ as $(c_1,\dots,c_n)$. Let $\cF=\{i_1,i_2\}$ be the indices
of the failed nodes, and let $\cR\subseteq [n]\backslash\{i_1,i_2\}$ be the set of $d$ helper nodes used in repair.
Our repair scheme is performed over the field 
\begin{equation}\label{eq:DFF}
F:=\mathbb{F}_p(\{\alpha_j:j\in [n]\setminus \{i_1,i_2\}\}).
\end{equation}
It is clear that $\mathbb{F}=F(\alpha_{i_1},\alpha_{i_2})$ and $[\mathbb{F}:F]=p_{i_1} p_{i_2}$.
As a consequence, $[\mathbb{K}:F]=s p_{i_1} p_{i_2}$.
Our strategy is as follows: 
\vspace{.1in}\begin{enumerate}
  \item[$(i)$] First repair node $c_{i_1}$ from the helper nodes in $\cR$. We show that this can be done by downloading 
$({s p_{i_1} p_{i_2}})/s_1$ symbols of $F$ from each of the helper nodes in $\cR$.
\vspace{.05in}  
  \item[$(ii)$] 
Then we use the helper nodes in $\cR$ together with the already repaired node $c_{i_1}$ to repair the node $c_{i_2}$, and we show that this can be done by downloading 
$\frac{s p_{i_1} p_{i_2}}{s_2}$ symbols of $F$ from each of the helper nodes in $\cR$.
\vspace{.05in} 
   \item[$(iii)$] We show that for each helper node in $\cR$, the two sets of downloaded symbols (for the repair of $c_{i_1}$ and $c_{i_2},$ respectively) have an overlap of size $p_{i_1} p_{i_2}$.
 \end{enumerate}  
\vspace{.1in} Therefore in total we need to download 
  \begin{align*}
  s_2p_{i_1} p_{i_2}&+s_1p_{i_1} p_{i_2}-p_{i_1} p_{i_2}\\
     &=2s_1p_{i_1} p_{i_2}\\
     &=\frac 2{s_2} s p_{i_1} p_{i_2}
  \end{align*}
   symbols of $F$ from each of the helper nodes. This forms a $2/(d+2-k)$ proportion of the node contents, and so the scheme achieves the cut-set bound \eqref{eq:cs1} with equality.


\vspace*{.1in} Proceeding with the implementation of the above plan, define the sets $W_{i_1}, W_{i_1}^{(1)}, W_{i_1}^{(2)}$
and $W_{i_2}, W_{i_2}^{(1)}, W_{i_2}^{(2)}$ as follows:
\begin{equation}\label{eq:wi}
\begin{aligned}
W_{i_1}^{(1)} := &
\Big\{ \beta^{u_1} \alpha_{i_1}^{u_1 + qs_1}  :
u_1=0,1,\dots,s_1-1; 
 q=0,1,\dots,\frac{p_{i_1}-1}{s_1}-1 \Big\}, \\
W_{i_1}^{(2)} := &
\Big\{  \alpha_{i_1}^{p_{i_1}-1} \sum_{u_1=0}^{s_1-1} \beta^{u_1} \Big\}, \\
W_{i_1} := & W_{i_1}^{(1)} \cup W_{i_1}^{(2)}; \\[.1in]
W_{i_2}^{(1)} := &
\Big\{ \beta^{u_2 s_1} \alpha_{i_2}^{u_2 + qs_2}  :
u_2=0,1,\dots,s_2-1; 
 q=0,1,\dots,\frac{p_{i_2}-1}{s_2}-1 \Big\}, \\
W_{i_2}^{(2)} := &
\Big\{   \alpha_{i_2}^{p_{i_2}-1}\sum_{u_2=0}^{s_2-1} \beta^{u_2 s_1} \Big\}, \\
W_{i_2} := & W_{i_2}^{(1)} \cup W_{i_2}^{(2)}.
\end{aligned}
\end{equation}

We further define two sets of elements
\begin{equation}\label{eq:defSi}
S_{i_1} := \bigcup_{u_2=0}^{s_2-1} \bigcup_{q_2=0}^{p_{i_2}-1}
\Big( \beta^{u_2 s_1} \alpha_{i_2}^{q_2} W_{i_1} \Big), \quad
S_{i_2} := \bigcup_{u_1=0}^{s_1-1} \bigcup_{q_1=0}^{p_{i_1}-1}
\Big( \beta^{u_1} \alpha_{i_1}^{q_1}  W_{i_2}\Big),
\end{equation}
where the product of an element $\alpha$ and a set $S$ is defined as the set $\alpha S = \{\gamma \alpha: \gamma \in S\}$.
It is clear that $|S_{i_1}|=s_2 p_{i_1} p_{i_2}$ and $|S_{i_2}|=s_1 p_{i_1} p_{i_2}$.

The theorem will follow from the next three lemmas.

\begin{lemma} \label{lem:S1}
Node $c_{i_1}$ can be repaired from the set of symbols 
$\{\trace_{\mathbb{K}/F}(\gamma v_jc_j):\gamma\in S_{i_1},j\in \cR\}$.
\end{lemma}

\begin{lemma} \label{lem:S2}
Node $c_{i_2}$ can be repaired from $c_{i_1}$ together with the set of symbols 
$\{\trace_{\mathbb{K}/F}(\gamma v_jc_j):\gamma\in S_{i_2},j\in \cR\}$.
\end{lemma}

For a vector space $V$ over a field $F$ and a set of vectors $A \subset V$, let $\spun_F(A)$ be the linear span of $A$ over $F$.

\begin{lemma} \label{lem:ints}
$$
\dim_F(\spun_F(S_{i_1}) \cap \spun_F(S_{i_2})) = p_{i_1} p_{i_2}.
$$
\end{lemma}

Let us first show that these three lemmas indeed imply Theorem~\ref{thm:2err}. 
On account of Lemmas \ref{lem:S1} and \ref{lem:S2} the sets of symbols
   $$
   D_j=\{\trace_{\mathbb{K}/F}(\gamma v_jc_j):\gamma\in S_{i_1}\cup S_{i_2}\}, \quad j\in \cR
   $$ 
suffice to find the values $c_{i_1}$ and $c_{i_2}.$
In their turn, the elements in the set $D_j, j\in \cR$ will be found once we download the elements in the set
$\{\trace_{\mathbb{K}/F}(\gamma v_jc_j):\gamma\in B\}$, where
 the elements in $B$ form a basis of 
$\spun_F(S_{i_1}) + \spun_F(S_{i_2})$ over $F$.
Therefore the number of symbols in $F$ that we need to download from each helper node is equal to the dimension of
$\spun_F(S_{i_1}) + \spun_F(S_{i_2})$ over $F$. We have
   \begin{equation}\label{eq:ios}
\dim_F(\spun_F(S_{i_1}) + \spun_F(S_{i_2}))= |S_{i_1}| + |S_{i_2}| - \dim_F(\spun_F(S_{i_1}) \cap \spun_F(S_{i_2})).
   \end{equation}
Using Lemma~\ref{lem:ints}, we now obtain
  $$
 \dim_F(\spun_F(S_{i_1}) + \spun_F(S_{i_2})) 
=  2 s_1 p_{i_1} p_{i_2} = \frac{2}{d+2-k} s p_{i_1} p_{i_2}.
  $$
Since $[\mathbb{K}:F]=s p_{i_1} p_{i_2}$, we conclude that the repair bandwidth of $\{c_{i_1},c_{i_2}\}$ from the helper nodes $\{c_j:j\in\cR\}$ indeed achieves the cut-set bound \eqref{eq:cs1}. 

Moreover, since the repair field of the pair $\{i_1,i_2\}$ is $\mathbb{F}_p(\{\alpha_j:j\in [n]\setminus \{i_1,i_2\}\})$, the largest common repair field for all possible pair of coordinates is ${\mathbb F}_p.$ This justifies the claim about the sub-packetization of our construction
made in \eqref{eq:2sp}.
\end{IEEEproof}

\vspace*{.1in}
Next we prove Lemmas~\ref{lem:S1}-\ref{lem:ints}.

\vspace*{.1in}{\em Proof of Lemma~\ref{lem:S1}:}
The proof of this lemma is an extension of the argument of Theorem \ref{thm1} (more on this in Remark \ref{remark2} in the end of this section).
Define the field 
\begin{equation}\label{eq:fi1}
F_{i_1}:=\mathbb{F}_p(\{\alpha_j:j\neq i_1\}).
\end{equation}
According to \eqref{eq:defFm}, we have
\begin{equation}\label{eq:rFi1}
\mathbb{F}= F_{i_1}(\alpha_{i_1}), \text{~and~} [\mathbb{F} : F_{i_1}]=p_{i_1}. 
\end{equation} 
Let $h_1(x)$ be the annihilator polynomial of the set $\{\alpha_j: j\in[n]\setminus (\cR\cup\{i_1\}) \}$, i.e., 
\begin{equation}\label{eq:defh1}
h_1(x)=\prod_{j\in[n]\setminus (\cR\cup\{i_1\})}(x-\alpha_j).
\end{equation}
As  remarked above \eqref{eq:grs}, the dual code of $\cC$ is $\cC^\bot=\text{\rm GRS}_{\mathbb{K}}(n,n-k,\Omega,v),$ where  $v=(v_1,\dots,v_n) \in (\mathbb{K}^*)^n.$ 
Clearly, $\deg(x^t h_1(x))\leq s_1-1+n-(d+1)<n-k$ for all $t=0,1,\dots,s_1-1,$ so for any such $t$ we have
\begin{equation}\label{eq:dualm}
(v_1\alpha_1^t h_1(\alpha_1),\dots,v_n \alpha_n^t h_1(\alpha_n))\in\cC^\bot.
\end{equation}
These $s_1$ dual codewords will  be used to recover the $i_1$-th coordinate. 
We define a set $T_{i_1}$ as follows:
\begin{equation}\label{eq:defT1}
T_{i_1} :=  \bigcup_{u_2=0}^{s_2-1} 
\Big( W_{i_1} \beta^{u_2 s_1} \Big).
\end{equation}
The elements in $T_{i_1}$ will also be used to recover the $i_1$-th coordinate.
Using \eqref{eq:defSi}, it is easy to verify the following relation:
\begin{equation}\label{eq:SRT}
S_{i_1}  = \bigcup_{q_2=0}^{p_{i_2}-1} T_{i_1} \alpha_{i_2}^{q_2}.
\end{equation}

Let $c=(c_1,\dots,c_n)\in \cC$ be a codeword, and let us construct a repair scheme for the coordinate (node) $c_i$ using
the values $\{c_j:j\in \cR\}$. Rewrite \eqref{eq:dualm} as follows:
$$
\sum_{j=1}^n  v_j\alpha_j^t h_1(\alpha_j) c_j =0, \quad  t=0,\dots,s_1-1.
$$
As an immediate consequence, for all $t=0,\dots,s_1-1$ and $\gamma \in T_{i_1},$ we have
\begin{equation}\label{eq:interm}
\sum_{j=1}^n \trace_{\mathbb{K}/F_{i_1}} (\gamma v_j\alpha_j^t h_1(\alpha_j) c_j) =0 .
\end{equation}
Let us write \eqref{eq:interm} in the following form:
\begin{equation}\label{eq:rcvm}
\begin{aligned}
\trace_{\mathbb{K}/F_{i_1}} (\gamma \alpha_{i_1}^t v_{i_1} h_1(\alpha_{i_1})c_{i_1}) & 
= - \sum_{j\neq i_1} \trace_{\mathbb{K}/F_{i_1}} (\gamma v_j \alpha_j^t h_1(\alpha_j)c_j) \\
& = - \sum_{j\in \cR} \trace_{\mathbb{K}/F_{i_1}} (\gamma v_j \alpha_j^t h_1(\alpha_j)c_j) \\
& = - \sum_{j\in \cR} \alpha_j^t h_1(\alpha_j) \trace_{\mathbb{K}/F_{i_1}} (\gamma v_j  c_j)
\text{~for all~} t=0,\dots,s_1-1 \text{~and all~} \gamma \in T_{i_1},
\end{aligned}
\end{equation}
where the second equality follows from \eqref{eq:defh1} and the third follows from the fact that the trace mapping $\trace_{\mathbb{K}/F_{i_1}}$ is $F_{i_1}$-linear, and that $\alpha_j\in F_{i_1}$
and $h_1(\alpha_j) \in F_{i_1}$ for all $j \neq i_1$.

Next we observe that the set $\{\gamma\alpha_{i_1}^t:  t=0,1,\dots,s_1-1 ;\, \gamma\in T_{i_1}\}$ of size $sp_{i_1}$ forms a basis of  $\mathbb{K}$ over  $F_{i_1}$ (see Prop.~\ref{prop:sumt} in Appendix~\ref{ap:PropT}). Since  $v_{i_1} h_1(\alpha_{i_1})\neq 0,$ the set
$\{\gamma\alpha_{i_1}^t v_{i_1} h_1(\alpha_{i_1}):  t=0,1,\dots,s_1-1 ;\,  \gamma\in T_{i_1}\}$  also forms a basis.
Therefore, the value of $c_{i_1}$ can be calculated from the set
  $$
  \{\trace_{\mathbb{K}/F_{i_1}} (\gamma \alpha_{i_1}^t v_{i_1} h_1(\alpha_{i_1})c_{i_1})
:t=0,1,\dots,s_1-1 ;\,  \gamma\in T_{i_1} \}.
 $$

Using \eqref{eq:rcvm}, we conclude that the value of $c_{i_1}$ can be calculated from
$\{\trace_{\mathbb{K}/F_{i_1}} (\gamma v_j  c_j):\gamma\in T_{i_1}, j\in\cR\}$.
To complete the proof of Lemma~\ref{lem:S1}, it suffices to show that the elements in the set
$\{\trace_{\mathbb{K}/F_{i_1}} (\gamma v_j  c_j):\gamma\in T_{i_1}, j\in\cR\}$
can be calculated from 
$\{\trace_{\mathbb{K}/F}(\gamma v_jc_j):\gamma\in S_{i_1},j\in \cR\}$. This is an immediate consequence of equation \eqref{eq:SRT}.
Indeed, observe that $F_{i_1}=F(\alpha_{i_2})$ and that $\{1,\alpha_{i_2},\dots,\alpha_{i_2}^{p_{i_2}-1}\}$ forms a basis of $F_{i_1}$ over $F$. Therefore, for every $\gamma\in T_{i_1}$ and every $j\in\cR$, the value of $\trace_{\mathbb{K}/F_{i_1}} (\gamma v_j  c_j)$ can be calculated from
$\{\trace_{F_{i_1}/F}(\trace_{\mathbb{K}/F_{i_1}} (\gamma v_j  c_j) \alpha_{i_2}^{q_2}): q_2=0,1,\dots,p_{i_2}-1\}$.
Observe that
$$
\trace_{F_{i_1}/F}(\trace_{\mathbb{K}/F_{i_1}} (\gamma v_j  c_j) \alpha_{i_2}^{q_2} )
= \trace_{F_{i_1}/F}(\trace_{\mathbb{K}/F_{i_1}} (\gamma v_j  c_j \alpha_{i_2}^{q_2}) )
= \trace_{\mathbb{K}/F} (\gamma v_j  c_j \alpha_{i_2}^{q_2}),
$$
where the first equality follows from the fact that $\alpha_{i_2}\in F_{i_1}$, and the second equality follows from \eqref{eq:trans}.
Therefore, for every $\gamma\in T_{i_1}$ and every $j\in\cR$, the value of $\trace_{\mathbb{K}/F_{i_1}} (\gamma v_j  c_j)$ can be calculated from
$\{\trace_{\mathbb{K}/F} (\gamma v_j  c_j \alpha_{i_2}^{q_2}):q_2=0,1,\dots,p_{i_2}-1\}
\subseteq \{\trace_{\mathbb{K}/F}(\gamma v_jc_j):\gamma\in S_{i_1},j\in \cR\}$, where the inclusion follows from \eqref{eq:SRT}.
Therefore we have shown that the elements in the set
$\{\trace_{\mathbb{K}/F_{i_1}} (\gamma v_j  c_j):\gamma\in T_{i_1}, j\in\cR\}$
can be calculated from 
$\{\trace_{\mathbb{K}/F}(\gamma v_jc_j):\gamma\in S_{i_1},j\in \cR\}$, and this completes the proof of Lemma~\ref{lem:S1}. \hfill $\blacksquare$

\vspace*{.1in}{\em Proof of Lemma~\ref{lem:S2}:}
Let $h_2(x)$ be the annihilator polynomial of the set $\{\alpha_j: j\in[n]\setminus (\cR\cup\{i_1,i_2\}) \}$, i.e., 
\begin{equation}\label{eq:defh2}
h_2(x)=\prod_{j\in[n]\setminus (\cR\cup\{i_1,i_2\})}(x-\alpha_j).
\end{equation}
Clearly, $\deg(x^t h_2(x))\leq s_2-1+n-(d+2)<n-k$ for all $t=0,1,\dots,s_2-1,$ so for any such $t$ we have
\begin{equation}\label{eq:dual2}
(v_1\alpha_1^t h_2(\alpha_1),\dots,v_n \alpha_n^t h_2(\alpha_n))\in\cC^\bot.
\end{equation}
These $s_2$ dual codewords will  be used to recover the $i_2$-th coordinate. 
Let us construct a repair scheme for the coordinate (node) $c_{i_2}$ using
the values $\{c_j:j\in \cR\cup\{i_1\}\}$. Rewrite \eqref{eq:dual2} as follows:
$$
\sum_{j=1}^n  v_j\alpha_j^t h_2(\alpha_j) c_j =0 \text{~for all~}  t=0,\dots,s_2-1.
$$
Computing the trace, we obtain
\begin{equation}\label{eq:inter2}
\sum_{j=1}^n \trace_{\mathbb{K}/F} (\gamma v_j\alpha_j^t h_2(\alpha_j) c_j) =0 \text{~for all~}  t=0,\dots,s_2-1 \text{~and all~} \gamma \in S_{i_2}.
\end{equation}
Let us write \eqref{eq:inter2} in the following form:
\begin{equation}\label{eq:rcv2}
\begin{aligned}
\trace_{\mathbb{K}/F} (\gamma \alpha_{i_2}^t v_{i_2} h_2(\alpha_{i_2})c_{i_2}) & 
= - \sum_{j\neq i_2} \trace_{\mathbb{K}/F} (\gamma v_j \alpha_j^t h_2(\alpha_j)c_j) \\
& = - \trace_{\mathbb{K}/F} (\gamma v_{i_1} \alpha_{i_1}^t h_2(\alpha_{i_1})c_{i_1})
- \sum_{j\in \cR} \trace_{\mathbb{K}/F} (\gamma v_j \alpha_j^t h_2(\alpha_j)c_j) \\
& = - \trace_{\mathbb{K}/F} (\gamma v_{i_1} \alpha_{i_1}^t h_2(\alpha_{i_1})c_{i_1})
 - \sum_{j\in \cR} \alpha_j^t h_2(\alpha_j) \trace_{\mathbb{K}/F} (\gamma v_j  c_j)\\
& \hspace*{0.4in} \text{~for all~} t=0,\dots,s_2-1 \text{~and all~} \gamma \in S_{i_2},
\end{aligned}
\end{equation}
where the second equality follows from \eqref{eq:defh2} and the third follows from the fact that the trace mapping $\trace_{\mathbb{K}/F}$ is $F$-linear, and that $\alpha_j\in F$ and $h_2(\alpha_j) \in F$ for all $j\in \cR$.

According to Prop.~\ref{prop:sumS} in Appendix~\ref{ap:PropT}, the set $\{\gamma\alpha_{i_2}^t:  t=0,1,\dots,s_2-1 ;\, \gamma\in S_{i_2}\}$ forms a basis of  $\mathbb{K}$ over  $F$
and so does the set  $\{\gamma\alpha_{i_2}^t v_{i_2} h_2(\alpha_{i_2}):  t=0,1,\dots,s_2-1 ;\,  \gamma\in S_{i_2}\}$ (recall that $v_{i_2} h_2(\alpha_{i_2})\neq 0$).
Hence the value of $c_{i_2}$ can be calculated from 
$\{\trace_{\mathbb{K}/F} (\gamma \alpha_{i_2}^t v_{i_2} h_2(\alpha_{i_2})c_{i_2})
:t=0,1,\dots,s_2-1 ;\,  \gamma\in S_{i_2} \}$.

Using \eqref{eq:rcv2}, we conclude that the value of $c_{i_2}$ can be calculated from
the value of $c_{i_1}$ and the values of elements in the set
$\{\trace_{\mathbb{K}/F} (\gamma v_j  c_j):\gamma\in S_{i_2}, j\in\cR\}$.
This completes the proof of Lemma~\ref{lem:S2}.
\hfill $\blacksquare$

\vspace*{.1in}{\em Proof of Lemma~\ref{lem:ints}:}
Using the cut-set bound on the left-hand side of Equation \eqref{eq:ios}, we obtain the inequality
    $$
\dim_F(\spun_F(S_{i_1}) \cap \spun_F(S_{i_2})) \le p_{i_1} p_{i_2}.
    $$
Let us prove that
     \begin{equation}\label{eq:isd}
\dim_F(\spun_F(S_{i_1}) \cap \spun_F(S_{i_2})) \ge p_{i_1} p_{i_2}.
     \end{equation}
To this end, we will find $p_{i_1} p_{i_2}$ elements in
$\spun_F(S_{i_1}) \cap \spun_F(S_{i_2})$ that are linearly independent over $F$.

Let us recall the definitions of $W_{i_1}$ and $W_{i_2}$ given in \eqref{eq:wi}.
Note that
$$
W_{i_2}  \subseteq \spun_F \Big(\bigcup_{u_2=0}^{s_2-1} \bigcup_{q_2=0}^{p_{i_2}-1} \{\beta^{u_2 s_1} \alpha_{i_2}^{q_2}\}\Big).
$$
Combining this with \eqref{eq:defSi}, we deduce that
$$
W_{i_1}\odot W_{i_2} \subseteq W_{i_1} \odot
\spun_F\Big(\bigcup_{u_2=0}^{s_2-1} \bigcup_{q_2=0}^{p_{i_2}-1} \{\beta^{u_2 s_1} \alpha_{i_2}^{q_2}\}\Big)
\subseteq \spun_F(S_{i_1}),
$$
where the product $\odot$ of sets $A_1$ and $A_2$ is defined as 
\begin{equation}\label{eq:dod}
A_1\odot A_2 
:=\{\gamma_1 \gamma_2:\gamma_1\in A_1, \gamma_2\in A_2\}.
\end{equation}
Similarly, we also have $W_{i_1}\odot W_{i_2} \subseteq \spun_F(S_{i_2})$, and therefore 
  \begin{equation}\label{eq:o.}
W_{i_1}\odot W_{i_2} \subseteq ( \spun_F(S_{i_1}) \cap \spun_F(S_{i_2}) ).
  \end{equation}
It is clear that 
$|W_{i_1}\odot W_{i_2}| = |W_{i_1}| |W_{i_2}| = p_{i_1} p_{i_2}$.
Moreover, for every $u\in\{0,1,\dots,s-1\}$, every $q_1\in\{0,1,\dots,p_{i_1}-1\}$ and every
$q_2\in\{0,1,\dots,p_{i_2}-1\}$, $\beta^u \alpha_{i_1}^{q_1} \alpha_{i_2}^{q_2}$
appears at most once\footnote{Such an element may be itself contained in $W_{i_1}\odot W_{i_2},$ or appear as a summand of an element in $W_{i_1}\odot W_{i_2}$} in $W_{i_1}\odot W_{i_2}$.
Since the elements in the set
$\{\beta^u \alpha_{i_1}^{q_1} \alpha_{i_2}^{q_2}: u=0,1,\dots,s-1; q_1=0,1,\dots,p_{i_1}-1;
q_2=0,1,\dots,p_{i_2}-1\}$ are linearly independent over $F$,
we deduce that all the elements in
$W_{i_1}\odot W_{i_2}$ are linearly independent over $F$.
Now \eqref{eq:isd} follows from \eqref{eq:o.}, and this completes the proof of Lemma~\ref{lem:ints}.
\hfill $\blacksquare$

\begin{remark}{\rm 
It is obvious from the proofs that the code construction in this section also has the $(1,d)$-optimal repair property and $(1,d+1)$-optimal repair property. In other words, the repair of any single erasure from any $d$ or $d+1$ helper nodes also achieves the cut-set bound.}
\end{remark}

\begin{remark} \label{remark2}
{\rm Let us point out some new ingredients in the repair of multiple erasures compared to the single-erasure case. These ideas will be used in the next section where we present a scheme for repairing an arbitrary number of erasures.

The first one appears in the proof of Lemma~\ref{lem:S1} whose proof consists of two parts: in the first part we show that $c_{i_1}$ can be calculated from $\{\trace_{\mathbb{K}/F_{i_1}} (\gamma v_j  c_j):\gamma\in T_{i_1}, j\in\cR\}$;
in the second part we show that the elements in the set $\{\trace_{\mathbb{K}/F_{i_1}} (\gamma v_j  c_j):\gamma\in T_{i_1}, j\in\cR\}$ can be calculated from $\{\trace_{\mathbb{K}/F}(\gamma v_jc_j):\gamma\in S_{i_1},j\in \cR\}$.
The proof of the first part is the same as the proof of Theorem~\ref{thm1}, and the new idea lies in the second part, where in particular we use transitivity of the trace mapping.

The other new ingredient is Lemma~\ref{lem:ints}, where we calculate the dimension of the intersection. Similar calculations also allow us to achieve the cut-set bound for the repair of more than two erasures in the next section. }
\end{remark}

\begin{remark}\label{remark3}
{\rm Finally, consider the full subfield lattice ordered by inclusion, starting with the field $\ff_p$ as the root and ending
with $\ff$ as the unique maximal element, i.e., the subset lattice of the $n$-set 
$\{\alpha_1,\alpha_2,\dots,\alpha_n\}$. In the above repair scheme we relied on subfields of the form $F$
(see \eqref{eq:DFF}), i.e., those that
contain all but two elements of this set. In a similar way, in our repair scheme for $h\ge 2$ erasures below we rely on subfields that contain $n-h$ of the $n$ elements of the set $\{\alpha_1,\alpha_2,\dots,\alpha_n\}$.
}
\end{remark}

\section{Universally achieving cut-set bound for any number of erasures}\label{Sect:uee}
In this section we present an explicit construction of $(n,k=n-r)$ RS codes with the universal $(h,d)$-optimal repair property for all $h\le r$ and all $k\le d \le n-h$ simultaneously. In other words, the constructed codes can repair any number of erasures from any set of helper nodes with repair bandwidth achieving the cut-set bound.
Even though the notation in this section is somewhat more involved than above, the main ideas are similar to the ideas used in the construction of RS codes with optimal repair for two erasures.

We again rely on the field tower introduced in Sec.~\ref{Sect:tower}, where in this case we take $s=r!.$
Our construction of codes with the universal $(h,d)$-optimal repair property relies on RS codes with evaluation points
$\alpha_1,\dots,\alpha_n.$ 
Specifically, the following is true:

\begin{theorem} \label{thm:ue} 
Let $k,n$ be any positive integers such that $k < n$ and let $p_i,i=1,2,\dots,n$ be the $i$th smallest prime that 
satisfies \eqref{eq:pmsm}. Let $\Omega=\{\alpha_1,\dots,\alpha_n\}$, where $\alpha_i,i=1,\dots,n$ is an element of degree $p_i$ over $\mathbb{F}_p.$
 The code $\cC:=\text{\rm RS}_{\mathbb{K}}(n,k,\Omega)$ achieves the cut-set bound for the repair of any number $h$ of failed nodes from any set of $d$ helper nodes provided that $h\le r$ and $k\le d \le n-h.$ 
In other words, $\cC$ has the universal $(h,d)$-optimal repair property for all $h$ and $d$ simultaneously.

The sub-packetization value of the code $\cC$ equals
   \begin{equation}\label{eq:sp}
   l=[\mathbb{K}:\mathbb{F}_p]=r!\prod_{i=1}^n p_i.
   \end{equation}
   For fixed $r$ and growing $n$ we have $l=e^{(1+o(1))n \log n}.$
 \end{theorem}

\begin{IEEEproof}
We write a codeword of $\cC$ as $(c_1,\dots,c_n)$.
Suppose that the number of failed nodes is $h$ and the number of helper nodes is $d$ for some $h\le r$ and some $k\le d \le n-h$.
Without loss of generality, we assume that
the indices of the failed nodes are $\cF=\{1,2,\dots,h\}$ and the indices of helper nodes
are $\cR=\{h+1,h+2,\dots,h+d\}$. 
Our repair scheme of these $h$ failed nodes is performed over the field
$$
F_{[h]}:=\mathbb{F}_p(\{\alpha_i:i\in [n]\setminus [h]\})
$$
(recall that $[h]:=\{1,2,\dots,h\}$; see also Remark~\ref{remark3}).
It is clear that $\mathbb{F}=F_{[h]}(\alpha_1,\alpha_2,\dots,\alpha_h)$ and $[\mathbb{F}:F_{[h]}]=\prod_{i=1}^h p_i$.
As a consequence, 
   \begin{equation}\label{eq:dim}
   [\mathbb{K}:F_{[h]}]=r! \prod_{i=1}^h p_i.
   \end{equation}
Our strategy is as follows: 
   \begin{enumerate}
  \item[$(i)$]
Begin with repairing node $c_1$ from the helper nodes in $\cR$. We show that this can be done by downloading 
$\frac{r! \prod_{i=1}^h p_i}{d+1-k}$ symbols of $F_{[h]}$ from each of the helper nodes in $\cR$.
\item[$(ii)$]
Then we use the helper nodes in $\cR$ together with the already repaired node $c_1$ to repair the node $c_2$, and we show that this can be done by downloading 
$\frac{r! \prod_{i=1}^h p_i}{d+2-k}$ symbols of $F_{[h]}$ from each of the helper nodes in $\cR$.
\item[$(iii)$]
We continue in this way until we use the helper nodes in $\cR$ together with the already repaired nodes $c_1,c_2,\dots,c_{h-1}$ to repair $c_h$.
\item[$(iv)$]
Finally we show that for each helper node in $\cR,$ the $h$ sets of downloaded symbols (for the repair of $c_1,c_2,\dots,c_h$ respectively) have overlaps, and that after removing the overlapping parts it suffices to download 
$\frac{h}{d+h-k}r! \prod_{i=1}^h p_i$ symbols of $F_{[h]}$ from each of the helper nodes, which achieves the cut-set bound \eqref{eq:cs1} with equality.
\end{enumerate}
\vspace*{.1in}

We introduce some notation before proceeding further.
Similarly to \eqref{eq:s12}, we define the following $h$ constants: for $i=1,2,\dots,h$, let
\begin{equation}\label{eq:sh}
s_i=d+i-k.
\end{equation}
Note that $s_i\le r$ for all $i\le h,$ and so $s_i|(p_i-1)$. It will also be convenient to have a notation for partial products of the numbers $s_i$. Namely, let
\begin{equation}\label{eq:tp}
t_1=1;\quad t_i=\prod_{j=1}^{i-1}s_j ,\; i=2,3,\dots,h+1
\end{equation}
and let 
    \begin{equation}\label{eq:sh+1}  
      s_{h+1}: = \frac{r!}{t_{h+1}}.
    \end{equation}
Observe the following simple facts:
\begin{align}
\Big\{\sum_{i=1}^h u_i t_i :
u_i=0,1,\dots,s_i-1; i=1,2,\dots,h\Big\} 
& =  \{0,1,2,\dots, t_{h+1} - 1\}, \nonumber \\ 
\Big\{\sum_{i=1}^{h+1} u_i t_i: 
u_i=0,1,\dots,s_i-1 \text{~for all~} i=1,2,\dots,h+1\Big\} 
& =  \{0,1,2,\dots, r! - 1\}. \label{eq:nsu}
\end{align}

For every $i\in[h]$, define three sets $W_i^{(1)},W_i^{(2)}$ and $W_i$ as follows:
\begin{equation}\label{eq:dwi}
\begin{aligned}
W_i^{(1)} & := \Big\{\beta^{u_i t_i} \alpha_i^{u_i + q s_i}:
u_i=0,1,\dots,s_i-1; q=0,1,\dots,\frac{p_i-1}{s_i}-1 \Big\}, \\
W_i^{(2)} & := \Big\{\sum_{u_i=0}^{s_i-1} \beta^{u_i t_i} \alpha_i^{p_i-1} \Big\}, \\
W_i & := W_i^{(1)} \cup W_i^{(2)}.
\end{aligned}
\end{equation}
We will also use the following notation. Let
  \begin{gather*}
  \mathbi{u}_{\sim i} := (u_1,u_2,\dots,u_{i-1},u_{i+1},\dots,u_{h+1})\\ 
\mathbi{q}_{\sim i} := (q_1,q_2,\dots,q_{i-1},q_{i+1}, \dots,q_h).
  \end{gather*}
For every $i=1,2,\dots,h$, let
\begin{align*}
U_{\sim i} & := \{\mathbi{u}_{\sim i} : u_j = 0,1,\dots,s_j-1 \text{~for all~}
j\in \{1,2,\dots,h+1\} \backslash\{i\}\}, \\
Q_{\sim i} & := \{\mathbi{q}_{\sim i} : q_j = 0,1,\dots,p_j-1 \text{~for all~}
j\in [h] \backslash\{i\}\}.
\end{align*}

Finally, define the set
$S_i, i=1,2,\dots,h$
\begin{equation}\label{eq:DSi}
\begin{aligned}
S_i  := \bigcup_{\mathbi{u}_{\sim i} \in U_{\sim i}}
\bigcup_{\mathbi{q}_{\sim i} \in Q_{\sim i}}
W_i \beta^{(\sum_{j=1;j\ne i}^{h+1} u_j t_j)} \prod_{j\in [h]\backslash \{i\}} \alpha_j^{q_j},
\end{aligned}
\end{equation}
 which we will use to characterize the symbols downloaded for repairing the $i$-th node.
Again let $\cC^\bot=\text{\rm GRS}_{\mathbb{K}}(n,n-k,\Omega,v)$ be the dual code of $\cC$ \eqref{eq:grs}, where the coefficients $v=(v_1,\dots,v_n) \in (\mathbb{K}^*)^n$ are nonzero.
The theorem will follow from the following two lemmas.

\begin{lemma} \label{lem:Sh}
Node $c_1$ can be repaired from the set of symbols
$\{\trace_{\mathbb{K}/F_{[h]}}(\gamma v_jc_j):\gamma\in S_1,j\in \cR\}$.
Node $c_i,i=2,3,\dots,h$ can be repaired from the values $c_1,c_2,\dots,c_{i-1}$ together with the set of symbols 
$\{\trace_{\mathbb{K}/F_{[h]}}(\gamma v_jc_j):\gamma\in S_i,j\in \cR\}$.
\end{lemma}

\begin{lemma} \label{lem:iSH}
   \begin{equation}\label{eq:iSH}
\dim_{F_{[h]}}\big(\spun_{F_{[h]}}(S_1) + \spun_{F_{[h]}}(S_2)
        + \ldots + \spun_{F_{[h]}}(S_h)\big) = \frac{h}{d+h-k}r! \prod_{i=1}^h p_i.
   \end{equation}
\end{lemma}

Once these lemmas are established, the proof of the theorem can be completed as follows. According to
Lemma~\ref{lem:Sh}, to recover the values of the nodes $c_1,c_2,\dots,c_h$ it suffices to know the elements in the set
$D_j=\{\trace_{\mathbb{K}/F_{[h]}}(\gamma v_jc_j):\gamma\in \cup_{i=1}^h S_i\}$ from each of the helper nodes $\{c_j:j\in\cR\}$. 
To calculate the values of elements in the set $D_j$, it suffices to download the elements in the set
$\{\trace_{\mathbb{K}/F_{[h]}}(\gamma v_jc_j):\gamma\in B\}$, where the elements in $B$ form a basis of 
$\spun_{F_{[h]}}(S_1) + \spun_{F_{[h]}}(S_2)
+ \ldots + \spun_{F_{[h]}}(S_h)$ over $F_{[h]}$.
By Lemma~\ref{lem:iSH}, the count of these elements equals $\frac{h}{d+h-k}r! \prod_{i=1}^h p_i.$
Combining this with \eqref{eq:dim}, we conclude that the repair of $c_1,c_2,\dots,c_h$ from the helper nodes $\{c_j:j\in\cR\}$ indeed achieves the cut-set bound \eqref{eq:cs1}.

Moreover, it is clear from the proof that the repair field of the $h$-tuple $\{i_1,i_2,\dots,i_h\}$ is $\mathbb{F}_p(\{\alpha_j:j\in [n]\setminus \{i_1,i_2,\dots,i_h\}\})$. Therefore the largest common repair field for all the possible $h$-tuples of coordinates is ${\mathbb F}_p.$ This justifies the claim about the sub-packetization of our construction
made in \eqref{eq:sp}.
\end{IEEEproof}

Next let us prove Lemmas~\ref{lem:Sh} and \ref{lem:iSH}.

\vspace*{.1in}{\em Proof of Lemma~\ref{lem:Sh}:}
For every $i\in[h]$, define a field
\begin{equation}\label{eq:f[i]}
F_{[i]}:=\mathbb{F}_p(\{\alpha_j:j\in[n]\backslash[i]\}).
\end{equation}
Fix $i\in[h]$ and let us prove the lemma for the repair of the $i$-th node.
Let $h_i(x)$ be the annihilator polynomial of the set $\{\alpha_j: j\in[n]\setminus (\cR\cup[i]) \}$, i.e., 
\begin{equation}\label{eq:hi}
h_i(x)=\prod_{j\in[n]\setminus (\cR\cup[i])}(x-\alpha_j).
\end{equation}
Clearly, $\deg(x^t h_i(x))\leq s_i-1+n-(d+i)<n-k$ for all $t=0,1,\dots,s_i-1,$ so for any such $t$ we have
\begin{equation}\label{eq:mdi}
(v_1\alpha_1^t h_i(\alpha_1),\dots,v_n \alpha_n^t h_i(\alpha_n))\in\cC^\bot.
\end{equation}
These $s_i$ dual codewords will  be used to recover the $i$-th coordinate. 
Further, define a set $T_i$ whose elements will also be used to recover the $i$th coordinate:
\begin{equation}\label{eq:dTi}
T_i   :=  
\bigcup_{\mathbi{u}_{\sim i} \in U_{\sim i}}
\bigcup_{q_1 = 0}^{p_1-1}
\bigcup_{q_2 = 0}^{p_2-1}
\ldots
\bigcup_{q_{i-1} = 0}^{p_{i-1}-1}
\Big(W_i \beta^{(\sum_{j=1;j\ne i}^{h+1} u_j t_j)} \prod_{1\le j<i} \alpha_j^{q_j} \Big).
\end{equation}
It is easy to verify the following relation:
\begin{equation}\label{eq:sti}
S_i   = \bigcup_{q_{i+1} = 0}^{p_{i+1}-1}
\bigcup_{q_{i+2} = 0}^{p_{i+2}-1}
\dots
\bigcup_{q_h = 0}^{p_h-1}
T_i  \prod_{i< j \le h} \alpha_j^{q_j}.
\end{equation}

Let $c=(c_1,\dots,c_n)\in \cC$ be a codeword, and let us construct a repair scheme for the coordinate (node) $c_i$ using
the values $\{c_j:j\in \cR\cup\{1,2,\dots,i-1\}\}$. Rewrite \eqref{eq:mdi} as follows:
$$
\sum_{j=1}^n  v_j\alpha_j^t h_i(\alpha_j) c_j =0 \text{~for all~}  t=0,1,\dots,s_i-1.
$$
Computing the trace, we obtain
\begin{equation}\label{eq:its}
\sum_{j=1}^n \trace_{\mathbb{K}/F_{[i]}} (\gamma v_j\alpha_j^t h_i(\alpha_j) c_j) =0 \text{~for all~}  t=0,\dots,s_i-1 \text{~and all~} \gamma \in T_i.
\end{equation}
Let us write \eqref{eq:its} in the following form:
\begin{equation}\label{eq:tig}
\begin{aligned}
\trace_{\mathbb{K}/F_{[i]}} (\gamma \alpha_i^t v_i h_i(\alpha_i)c_i) & 
= - \sum_{j\neq i} \trace_{\mathbb{K}/F_{[i]}} (\gamma v_j \alpha_j^t h_i(\alpha_j)c_j) \\
& = - \sum_{j=1}^{i-1} \trace_{\mathbb{K}/F_{[i]}} (\gamma v_j \alpha_j^t h_i(\alpha_j)c_j)
- \sum_{j\in \cR} \trace_{\mathbb{K}/F_{[i]}} (\gamma v_j \alpha_j^t h_i(\alpha_j)c_j) \\
& = - \sum_{j=1}^{i-1} \trace_{\mathbb{K}/F_{[i]}} (\gamma v_j \alpha_j^t h_i(\alpha_j)c_j)
- \sum_{j\in \cR} \alpha_j^t h_i(\alpha_j) \trace_{\mathbb{K}/F_{[i]}} (\gamma v_j  c_j) \\
& \hspace*{1.5in} \text{~for all~} t=0,\dots,s_i-1 \text{~and all~} \gamma \in T_i,
\end{aligned}
\end{equation}
where the second equality follows from \eqref{eq:hi} and the third follows from the fact that the trace mapping $\trace_{\mathbb{K}/F_{[i]}}$ is $F_{[i]}$-linear, and that $\alpha_j\in F_{[i]}$
and $h_i(\alpha_j) \in F_{[i]}$ for all $j \in \cR$.

\vspace*{.1in}According to Prop.~\ref{prop:mt} in Appendix~\ref{ap:PropT}, the set $\{\gamma\alpha_i^t:  t=0,1,\dots,s_i-1 ;\, \gamma\in T_i\}$ forms a basis\footnote{Note that the size of this set is $s_i|T_i|=(\prod_{j=1}^i p_j)( \prod_{m=1}^{h+1}s_m)$ 
which equals the extension degree $[{\mathbb K}:F_{[i]}]$ because of our definition of $s_{h+1}$ in \eqref{eq:sh+1}.} of  $\mathbb{K}$ over  $F_{[i]}$
and so does the set  $\{\gamma\alpha_i^t v_i h_i(\alpha_i):  t=0,1,\dots,s_i-1 ;\,  \gamma\in T_i\}$ (recall again that $v_i h_i(\alpha_i)\neq 0$).
Hence the value of $c_i$ can be calculated from 
$\{\trace_{\mathbb{K}/F_{[i]}} (\gamma \alpha_i^t v_i h_i(\alpha_i)c_i)
:t=0,1,\dots,s_i-1 ;\,  \gamma\in T_i \}$.

Using \eqref{eq:tig}, we conclude that the value of $c_i$ can be calculated from
the values of $c_1,c_2,\dots,c_{i-1}$ and the values of elements in the set
$\{\trace_{\mathbb{K}/F_{[i]}} (\gamma v_j  c_j):\gamma\in T_i, j\in\cR\}$.
The proof will be complete once we show that these elements can be found
from the elements in the set $\{\trace_{\mathbb{K}/F_{[h]}}(\gamma v_jc_j):\gamma\in S_i,j\in \cR\}$. This is an immediate consequence of \eqref{eq:trans} and equation \eqref{eq:sti}.
Indeed, observe that $F_{[i]}=F_{[h]}(\alpha_{i+1},\alpha_{i+2},\dots,\alpha_h),$ and that 
$\{\prod_{i< m \le h} \alpha_m^{q_m}: q_m=0,1,\dots,p_m-1, \forall i<m\le h\}$
 forms a basis of $F_{[i]}$ over $F_{[h]}$. Therefore, for every $\gamma\in T_i$ and every $j\in\cR$, the value of $\trace_{\mathbb{K}/F_{[i]}} (\gamma v_j  c_j)$ can be calculated from
    $$
    \Big\{\trace_{F_{[i]}/F_{[h]}}\Big(\trace_{\mathbb{K}/F_{[i]}} (\gamma v_j  c_j) \prod_{i< m \le h} \alpha_m^{q_m}\Big): q_m=0,1,\dots,p_m-1, \forall i<m\le h\Big\}.
    $$
Involving transitivity of the trace \eqref{eq:trans}, we see that
     \begin{align*}
\trace_{F_{[i]}/F_{[h]}}\Big(\trace_{\mathbb{K}/F_{[i]}} (\gamma v_j  c_j) \prod_{i< m \le h} \alpha_m^{q_m}\Big)
&= \trace_{F_{[i]}/F_{[h]}}(\trace_{\mathbb{K}/F_{[i]}} (\gamma v_j  c_j \prod_{i< m \le h} \alpha_m^{q_m} ) )\\
&= \trace_{\mathbb{K}/F_{[h]}} (\gamma v_j  c_j \prod_{i< m \le h} \alpha_m^{q_m} ),
     \end{align*}
where the first equality follows from the fact that $\alpha_m\in F_{[i]}$ for all $m>i.$
Therefore, for every $\gamma\in T_i$ and every $j\in\cR$, the value of $\trace_{\mathbb{K}/F_{[i]}} (\gamma v_j  c_j)$ can be calculated from
$$
\Big\{\trace_{\mathbb{K}/F_{[h]}} \Big(\gamma v_j  c_j \prod_{i< m \le h} \alpha_m^{q_m} \Big): q_m=0,1,\dots,p_m-1, \forall i<m\le h\Big\}
\subseteq \Big\{\trace_{\mathbb{K}/F_{[h]}}(\gamma v_jc_j):\gamma\in S_i,j\in \cR\Big\},
$$
 where the inclusion follows from \eqref{eq:sti}.
This establishes the needed fact, namely, that the elements in the set
$\{\trace_{\mathbb{K}/F_{[i]}} (\gamma v_j  c_j):\gamma\in T_i, j\in\cR\}$
can be calculated from 
$\{\trace_{\mathbb{K}/F_{[h]}}(\gamma v_jc_j):\gamma\in S_i,j\in \cR\}$, and completes the proof of Lemma~\ref{lem:Sh}. \hfill $\blacksquare$

The proof of Lemma \ref{lem:iSH} is given in Appendix \ref{ap:lem:iSH}.

\vspace*{.1in}

\section{Asymptotically optimal single-node repair RS codes with $l=r^n$}\label{RS}
In this section we construct a family of RS codes that do not achieve the cut-set bound, but approach it as the
block length $n$ becomes large. This result is accomplished by coupling the linear repair scheme of 
\cite{Guruswami16} with the $r$-ary expansion idea of
\cite{Cadambe11,Tamo13}.  
Suppose that $n$ and $k$ are arbitrary fixed numbers. Let $F$ be a finite field and let $h(x)\in F[x]$ be a degree 
$l$ irreducible polynomial over $F,$ where $l=r^n,r=n-k.$ Let $\beta$ be a root of $h(x)$ and set the symbol field to be $E=F(\beta),$ 
i.e., the field generated by $\beta$ over $F.$ Clearly $\{1,\beta,\beta^2,\dots,\beta^{l-1}\}$ is a basis for $E$ over 
$F.$ Choose the set of evaluation points to be $\Omega=\{\beta^{r^0},\beta^{r^1},\dots,\beta^{r^{n-1}}\}.$

\begin{theorem}\label{thm:powers}.
The repair bandwidth of the code $\text{RS}(n,k,\Omega)$ over $F$ is less than $l\frac{n+1}{n-k}.$
\end{theorem}
\begin{IEEEproof}
We need to show that for every $i\in[n],$ we can find polynomials $f_{i,j}$ with $\deg(f_{i,j})< r, j=1,\dots,l$ 
such that $f_{i,1}(\beta^{r^{i-1}}),\dots,f_{i,l}(\beta^{r^{i-1}})$ form a basis for $E$ over $F$ and 
    $$
    \sum_{0\le t<n,t\neq i-1} {\dim}_F(\{f_{i,j}(\beta^{r^t})\}_{j\in[l]})< \frac{l(n+1)}{n-k}.
    $$ 
    For $a=0,1,\dots,l-1,$ write its $r$-ary expansion as $a=(a_n,a_{n-1},\dots,a_1),$ where $a_i$ is the $i$-th digit from the right.
Define the set of $l$ polynomials $\{f_{i,j}\}_{j\in[l]}=\{\beta^a x^s:a_i=0,s=0,1,\dots,r-1\}.$
 
It is easy to verify that 
$$\{f_{i,j}(\beta^{r^{i-1}}): j\in[l]\}=\{1,\beta,\beta^2,\dots,\beta^{l-1}\}
$$ 
(as sets), so the elements 
$\{f_{i,j}(\beta^{r^{i-1}})\}_{j\in[l]}$ form a basis for $E$ over $F.$ When $t< i-1,$ we have 
\begin{align*}
\{f_{i,j}&(\beta^{r^t})\}_{j\in[l]}=\{\beta^a:a_i=0\}\bigcup\\
&\Big(\bigcup_{u=0}^{r-2}\{\beta^a:a_i=1,a_{i-1}=\dots=a_{t+2}=0,a_{t+1}=u\}\Big).
\end{align*}
Thus $\dim_F(\{f_{i,j}(\beta^{r^t})\}_{j\in[l]})\le \frac{l}{r}+(r-1)\frac{l}{r^{i-t}}$ if $t<i-1.$
When $t>i-1,$ we have
\begin{align*}
\{f_{i,j}&(\beta^{r^t})\}_{j\in[l]}=\{\beta^a:a_i=0\}\bigcup\\
&\Big(\bigcup_{u=0}^{r-2}\{\beta^{l+a}:a_n=\dots=a_{t+2}=0,a_{t+1}=u,a_i=0\}\Big).
\end{align*}
Thus ${\dim}_F(\{f_{i,j}(\beta^{r^t})\}_{j\in[l]})\le \frac{l}{r}+(r-1)\frac{l}{r^{n-t+1}}$ for $t>i-1.$
An upper bound on the sum of the dimensions is given by:
\begin{align*}
\sum_{0\le t<n,t\neq i-1}\!\!\!\! {\dim}_F(\{f_{i,j}(\beta^{r^t})\}_{j\in[l]})
&\le (n-1)\frac{l}{r}+(r-1)\sum_{t=0}^{i-2}\frac{l}{r^{i-t}}+(r-1)\sum_{t=i}^{n-1}\frac{l}{r^{n-t+1}} \\
&=l\Big(\frac{n-1}{r}+\frac{r^{i-1}-1}{r^i}+\frac{r^{n-i}-1}{r^{n-i+1}}\Big)\\
&< \,l\frac{n+1}{n-k}.
\end{align*}
The proof is complete.
\end{IEEEproof}
Since the optimal repair 
bandwidth for an $(n,k,l)$ MDS array code is $\frac{l(n-1)}{n-k}$, we conclude that when $n\to \infty,$ the ratio 
between the actual and the optimal repair bandwidth approaches $1$ (the corresponding quantity of the construction in \cite{Guruswami16} is about 1.5).
\remove{
\begin{remark}
In \cite{Guruswami16}, the authors gave a coset construction of a family of $2$-parity RS codes and a repair scheme which needs to download $(\frac{3}{2}+o(1))b_{\text{op}}$ sub-symbols to repair a single node failure when the code length goes to infinity, where $b_{\text{op}}$ is the optimal repair bandwidth of the MDS array codes with the same parameters. In contrast, if we fix the number of parity nodes to be $2,$ our construction and repair scheme above only needs to download $(1+o(1))b_{\text{op}}$ sub-symbols when the code length goes to infinity.
\end{remark}}
 
\section{Concluding remarks}
Let us point out some open problems related to the topic of this paper. One of them is establishing limits of repair of full-length
RS codes, i.e., taking the code length equal to the size $q$ of the symbol field. While shortened codes such as constructed above
can be optimally repaired, full-length codes cannot \cite{Guruswami16}. While \cite{Guruswami16, Dau16,Dau17,Bartan17} contain
some results along these lines, the full picture is far from being clear.

Switching to the topic of cooperative repair, note that it is possible to construct array MDS codes that achieve the
corresponding cut-set bound for the repair of any number of failed nodes \cite{YeBarg18}. At the same time, similar results
for RS codes are not yet available. Specifically, is it possible to modify the scheme in Sec.~\ref{sect:warmup-multiple} to attain
optimal cooperative repair of two erasures with RS codes?

The repair scheme of \cite{Guruswami16} was recently extended in \cite{JinLuoXing17} to general codes on algebraic curves. It is natural to address the question of extending the constructions of this paper to reduce the repair bandwidth of codes on curves (for instance, Hermitian codes) compared to the general results in \cite{JinLuoXing17}.

Finally, while optimal repair requires large sub-packetization $l$, stepping away from the cut-set bound enables one
to attain a very significant decrease of the node size \cite{Rawat2018}. It would be interesting to address this question 
for RS codes both for the full-length case and for the shortened version of this paper.

\appendices
\section{Proof of Claim \ref{claim1}}\label{ap:claim}
Consider the repair of the $j$-th node of the code $\cC$ for some $j\in \{1,2,\dots,k-1\}$.
Since $\cC$ {can be viewed as} an $(n,k,n-1,l)$ MSR code with a linear repair scheme over $\mathbb{F}_q$, node
$c_j$ can be repaired by downloading $(n-1)l/r$ symbols of $\mathbb{F}_q$ from all the remaining nodes $\{c_i:i\in[n]\setminus\{j\}\},$ where $r=n-k.$ Therefore by  Theorem~\ref{Thm:Guru}, there exist $l$ codewords 
$$
(c_{t,1},c_{t,2},\dots,c_{t,n}) \in \cC^\perp, t=1,2,\dots,l
$$ 
such that
\begin{align}
\dim_{\mathbb{F}_q}(c_{1,j},c_{2,j},\dots,c_{l,j})&= l, \text{ and }  \label{eq:21}\\
\sum_{i\neq j} \dim_{\mathbb{F}_q}(c_{1,i},c_{2,i},\dots,c_{l,i})&=\frac{(n-1)l}{r}. \label{eq:212}
\end{align}
Since $H$ is a generator matrix of $\cC^\perp$, for each $t=1,2,\dots,l$ there is a column vector 
$b_t \in E^r$ such that $(c_{t,1},c_{t,2},\dots,c_{t,n})=b_t^T H$.
We define an $l \times r$ matrix $B$ over the field $E$ as $B=[b_1 b_2 \dots b_l]^T$.
We claim that the $\mathbb{F}_q$-rank of the row space of $B$ is $l$. Indeed, assume the contrary, then there  exists a nonzero vector $w\in \mathbb{F}_q^l$ such that $wB=0$. 
Therefore,
$$
wBH=w
\left[ \begin{array}{cccc}
c_{1,1} & c_{1,2} & \dots & c_{1,n} \\
c_{2,1} & c_{2,2} & \dots & c_{2,n} \\
\vdots & \vdots & \vdots & \vdots \\
c_{l,1} & c_{l,2} & \dots & c_{l,n} \\
\end{array} \right]=0.
$$ 
This implies that $w(c_{1,j},c_{2,j},\dots,c_{l,j})^T=0,$
contradicting \eqref{eq:21}. 
Thus we conclude that $B$ has $l$ linearly independent rows over $\mathbb{F}_q$.

Now we want to show that there exists an $l \times l$ invertible matrix $A$ over $\mathbb{F}_q$ such that the matrix $AB$ is an $r\times r$ block-diagonal matrix $\Diag(a_1,\dots,a_r)$, where each block $a_i$ is formed of a column vector of length $\frac{l}{r}$.
In other words, by performing elementary row operations over $\mathbb{F}_q$, $B$ can be transformed into an $r\times r$ block-diagonal matrix $\Diag(a_1,\dots,a_r)$.
Indeed, for $i\in[n]$, let  $h_i$ be the $i$-th column of the matrix $H$, and define 
$$
t_i=\dim_{\mathbb{F}_q}(Bh_i)
= \dim_{\mathbb{F}_q}(c_{1,i},c_{2,i},\dots,c_{l,i}).
$$
By \eqref{eq:212}, we have
\begin{equation} \label{eq:sumt}
\sum_{i\neq j}^n t_i =\frac{(n-1)l}{r}. 
\end{equation}

Since $H$ generates an $(n,r)$ MDS code, for any subset of indices $\cJ \subseteq [n]$ of size $|\cJ|=r$, the matrix $H_{\cJ}$ is of full rank. 
Therefore, the $l\times r$ matrix $BH_{\cJ}$ satisfies the conditions
    \begin{equation}
l=\dim(\cS_{{\mathbb F}_q}(B))=\dim(\cS_{{\mathbb F}_q}(BH_{\cJ}))\leq \sum_{i\in \cJ}\dim_{\mathbb{F}_q}(Bh_i),
\label{eq:22}
\end{equation} 
where the last inequality follows from Lemma~\ref{lem:use}.
Summing both sides of  \eqref{eq:22} over all subsets $\cJ\subseteq [n]\backslash\{j\}$ of size $|\cJ|=r$, we obtain that
\begin{equation}\label{eq:sand}
\begin{aligned}
l \binom{n-1}{r}&\leq \sum_{\substack{\cJ \subseteq [n]\backslash\{j\} \\|\cJ|=r}}\sum_{i\in \cJ}\dim_{\mathbb{F}_q}(Bh_i) \\
&=\binom{n-2}{r-1}\sum_{i\neq j}t_i  \\
&\overset{\eqref{eq:sumt}}{=} \binom{n-2}{r-1}\frac{(n-1) l}{r}  \\
&=l\binom{n-1}{r},
\end{aligned}
\end{equation}
 This implies that the inequality above is in fact an equality, and therefore on account of \eqref{eq:22}, for every subset $\cJ\subseteq [n]\backslash\{j\},|\cJ|=r$ we have
\begin{equation}
l =\sum_{i\in \cJ}\dim_{\mathbb{F}_q}(Bh_i)=\sum_{i\in \cJ}t_i.
\label{eq:222}
\end{equation}
From \eqref{eq:222} we obtain that for all $i\in[n]\setminus\{j\}$ 
\begin{equation}\label{eq:sing}
\dim_{\mathbb{F}_q}(Bh_i)=t_i=l/r.
\end{equation}
Moreover, since \eqref{eq:22} holds with equality, we can use the second part of Lemma~\ref{lem:use} to claim that, for 
$\cJ\subseteq [n]\backslash\{j\}$ of size $|\cJ|\le r,$
    \begin{equation}\label{eq:tem}
\dim(\cS_{\mathbb{F}_q}(BH_{\cJ}))
= \sum_{i\in \cJ}\dim_{\mathbb{F}_q}(Bh_i)
= \frac{|\cJ|l}{r}. 
    \end{equation}
Let us take $\cJ$ to be a subset of $\{k+1,k+2,\dots,n\}.$ 
Since the last $r$ columns of $H$ form an identity matrix, \eqref{eq:tem} becomes
    \begin{equation}\label{eq:subm}
\dim(\cS_{\mathbb{F}_q}(B_{\cJ}))
= \frac{|\cJ|l}{r} 
\text{~for all~} \cJ\subseteq [r] \text{~with size~} |\cJ|\le r.
   \end{equation}

Now we are ready to prove that by performing elementary row operations over $\mathbb{F}_q$, $B$ can be transformed into an $r\times r$ block diagonal matrix $\Diag(a_1,\dots,a_r)$, where each block $a_i$ is a single column vector of length $\frac{l}{r}$. 
We proceed by induction.
More specifically, we prove that for $i=1,2,\dots,r$, we can use elementary row operations over $\mathbb{F}_q$ to transform the first $i$ columns of $B$ into the following form:
$$
\left[
\begin{array}{cccc}
a_1 & 0 & \dots & 0 \\
0 & a_2 & \dots & 0 \\
\vdots & \vdots & \vdots & \vdots \\
0 & 0 & \dots & a_i \\
\mathbf{0} & \mathbf{0} & \dots & \mathbf{0}
\end{array}
\right],
$$
where each $\mathbf{0}$ in the last row of the above matrix is a column vector of length $l(1-\frac{i}{r})$.

Let $i=1.$ According to \eqref{eq:subm}, each column of $B$ has dimension $l/r$ over $\mathbb{F}_q$. Thus the induction base holds trivially.
Now assume that there is an $l \times l$ invertible matrix $A$ over $\mathbb{F}_q$ such that
$$
A B_{[i-1]}=
\left[
\begin{array}{cccc}
a_1 & 0 & \dots & 0 \\
0 & a_2 & \dots & 0 \\
\vdots & \vdots & \vdots & \vdots \\
0 & 0 & \dots & a_{i-1} \\
\mathbf{0} & \mathbf{0} & \dots & \mathbf{0}
\end{array}
\right],
$$
where each $\mathbf{0}$ in the last row of this matrix is a column vector of length $l(1-\frac{i-1}{r})$.
Let us write the $i$-th column of $AB$ as $(v_1,v_2,\dots,v_l)^T$. 
Since each column of $B$ has dimension $l/r$ over $\mathbb{F}_q$,
$(v_1,v_2,\dots,v_l)^T$ also has dimension $l/r$ over $\mathbb{F}_q$.
Since the last $l(1-\frac{i-1}{r})$ rows of the matrix $A B_{[i-1]}$ are all zero, we can easily deduce that 
   $$
\dim(\cS_{\mathbb{F}_q}(AB_{[i]})) \le \frac{i-1}{r}l + 
\dim_{\mathbb{F}_q}(v_{(i-1)l/r+1}, v_{(i-1)l/r+2}, \dots, v_l).
   $$
By \eqref{eq:subm}, $\dim(\cS_{\mathbb{F}_q}(AB_{[i]}))=\dim(\cS_{\mathbb{F}_q}(B_{[i]}))=\frac{il}{r}$. 
As a result, 
$$
\dim_{\mathbb{F}_q}(v_{(i-1)l/r+1}, v_{(i-1)l/r+2}, \dots, v_l) \ge l/r = 
\dim_{\mathbb{F}_q}(v_1,v_2, \dots, v_l).
$$
In other words, $(v_{(i-1)l/r+1}, v_{(i-1)l/r+2}, \dots, v_l)$ contains a basis of the set $(v_1,v_2, \dots, v_l)$ over $\mathbb{F}_q$.
This implies that we can use elementary row operations on the matrix $AB$ to eliminate all the nonzero entries
$v_m$ for $m\le (i-1)l/r$, and thus obtain the desired block-diagonal structure for the first $i$ columns.
This establishes the induction step.

We conclude that there exists an $l \times l$ invertible matrix $A$ over $\mathbb{F}_q$ such that  $AB=\Diag(a_1,\dots,a_r)$, where each block $a_i$ is a single column vector of length $\frac{l}{r}$.
For $u\in[r]$, let $A_u$ be the vector space spanned by the entries of $a_u$ over $\mathbb{F}_q$.
According to \eqref{eq:sing}, for all $i\in[n]\setminus\{j\}$
$$
\dim_{\mathbb{F}_q}(ABh_i)=\dim_{\mathbb{F}_q}(Bh_i)=l/r.
$$
Since 
\begin{align*}
\dim_{\mathbb{F}_q}(ABh_i)&=\dim_{\mathbb{F}_q}(\Diag(a_1,\dots,a_r) h_i)\\
&=
\dim_{\mathbb{F}_q}(A_1h_{1,i}+\dots+ A_r h_{r,i}), \quad i=1,2,\dots,n,
\end{align*}
for all $i\in[n]\setminus\{j\}$ we have 
$$
\dim_{\mathbb{F}_q}(A_1h_{1,i}+\dots+ A_r h_{r,i}) =l/r.
$$
Since each column of $B$ has dimension $l/r$ over $\mathbb{F}_q$,
$A_u$ also has dimension $l/r$ over $\mathbb{F}_q$ for every $u\in[r]$.
Recall that $h_{u,i}\neq 0$ for all $u\in[r]$ and all $i\in[k]$.
Thus
$$
\dim_{\mathbb{F}_q}(A_u h_{u,i})=l/r
= \dim_{\mathbb{F}_q}(A_1h_{1,i}+\dots+ A_r h_{r,i})
$$
for all  $u=1,\dots,r$ and $i\in[k]\setminus\{j\}.$
Therefore,
$$
A_1 h_{1,i}=A_2 h_{2,i} = \dots = A_r h_{r,i} \text{~and all~} i\in[k]\setminus\{j\}.
$$
Since $h_{1,i}=1$ for all $i=1,2,\dots,k$, we have
\begin{equation}\label{eq:eqspa}
A_2 h_{2,i} = A_1 \text{~for all~} i\in[k]\setminus\{j\}.
\end{equation}
Equivalently,
$$
A_2 \alpha_i = A_2 \text{~for all~} i\in\{1,2,\dots,k-1\} \setminus\{j\}.
$$
By definition $A_2$ is a vector space over $\mathbb{F}_q$, so 
\begin{equation}\label{eq:stab}
A_2 \gamma = A_2 \text{~for all~} 
\gamma \in \mathbb{F}_q(\{\alpha_i: i\in\{1,2,\dots,k-1\} \setminus\{j\}\}).
\end{equation}
On the other hand,
\begin{equation}\label{eq:cond1}
\begin{aligned}
\dim_{\mathbb{F}_q}(A_1h_{1,j}+\dots+ A_r h_{r,j}) & =\dim_{\mathbb{F}_q}(\Diag(a_1,\dots,a_r) h_j)=
\dim_{\mathbb{F}_q}(ABh_j)\\
& = \dim_{\mathbb{F}_q}(Bh_j) = 
\dim_{\mathbb{F}_q}\{c_{1,j},c_{2,j},\dots,c_{l,j}\}= l,
\end{aligned}
\end{equation}
while
\begin{equation}\label{eq:cond2}
\dim_{\mathbb{F}_q}(A_u h_{u,j})=l/r, \quad  u=1,2,\dots,r.
\end{equation}
Equations \eqref{eq:cond1} and \eqref{eq:cond2} together imply that the vector spaces $A_1 h_{1,j}, A_2 h_{2,j}, \dots, A_r h_{r,j}$ are pairwise disjoint. In particular, $A_1 \cap A_2 h_{2,j} = \{0\}.$ On account of
\eqref{eq:eqspa}, we therefore have $A_2 h_{2,k} \cap A_2 h_{2,j} = \{0\}.$
This implies that $A_2 \alpha_j \neq A_2$. By \eqref{eq:stab}, we conclude that
$\alpha_j \notin \mathbb{F}_q(\{\alpha_i: i\in\{1,2,\dots,k-1\} \setminus\{j\}\})$.
This completes the proof of Claim \ref{claim1}.

\section{Technical proofs}\label{ap:PropT}
\begin{proposition}\label{prop:sumt}
For the set $T_{i_1}$ defined in \eqref{eq:defT1}, we have
$$
  \spun_{F_{i_1}}(T_{i_1}) + \spun_{F_{i_1}} (T_{i_1} \alpha_{i_1}) +\dots + 
\spun_{F_{i_1}} (T_{i_1} \alpha_{i_1}^{s_1-1} )=\mathbb{K},
$$
where $S \alpha := \{\gamma \alpha: \gamma \in S\}$,
and the operation $+$ is the Minkowski sum of sets, $T_1 + T_2 := \{\gamma_1+\gamma_2:\gamma_1\in T_1, \gamma_2\in T_2 \}.$ 
\end{proposition}
\begin{IEEEproof}
To establish the proposition, we will prove the following claim:
\begin{equation}\label{eq:cW}
  \spun_{F_{i_1}}(W_{i_1}) + \spun_{F_{i_1}} (W_{i_1} \alpha_{i_1}) +\dots + 
\spun_{F_{i_1}} (W_{i_1} \alpha_{i_1}^{s_1-1} )=\oplus_{u_1=0}^{s_1-1}\beta^{u_1}\mathbb{F}.
\end{equation}
Note that \eqref{eq:defT1} and \eqref{eq:cW} together imply that
\begin{align*}
  \spun_{F_{i_1}}(T_{i_1}) &+ \spun_{F_{i_1}} (T_{i_1} \alpha_{i_1}) +\dots + 
\spun_{F_{i_1}} (T_{i_1} \alpha_{i_1}^{s_1-1} ) \\
&=  \oplus_{u_1=0}^{s_1-1}\oplus_{u_2=0}^{s_2-1}\beta^{u_1+u_2s_1}\mathbb{F}\\
&= \oplus_{u=0}^{s-1} \beta^u \mathbb{F}\\
&= \mathbb{K},
\end{align*}
where the last equality follows from the fact that, on account of \eqref{eq:bbk}, the set $1,\beta,\dots,\beta^{s-1}$ forms a basis of $\mathbb{K}$ over $\mathbb{F}.$ Therefore the proposition indeed follows from \eqref{eq:cW}. 

Now we are left to prove \eqref{eq:cW}. Our arguments follow the proof of Lemma \ref{lem1}.

Let 
    $$
K:=\spun_{F_{i_1}}(W_{i_1}) + \spun_{F_{i_1}} (W_{i_1} \alpha_{i_1}) +\dots + 
\spun_{F_{i_1}} (W_{i_1} \alpha_{i_1}^{s_1-1} ).
   $$
Let us prove that $K=\oplus_{u_1=0}^{s_1-1}\beta^{u_1}\mathbb{F}.$ Clearly $K$ is a vector space over $F_{i_1}$, and by \eqref{eq:rFi1} we have $K\subseteq \oplus_{u_1=0}^{s_1-1}\beta^{u_1}\mathbb{F}$. 
Let us show the reverse inclusion, namely that $\oplus_{u_1=0}^{s_1-1}\beta^{u_1}\mathbb{F}\subseteq K$. More specifically, we will show that 
$\beta^{u_1}\mathbb{F} \subseteq K$ for all $u_1=0,1,\dots,s_1-1.$ 

We use induction on $u_1.$
For the induction base, let $u_1=0$, and let us show that the field $\mathbb{F}$ defined in \eqref{eq:defFm} is contained in $K$.
In this case, we have $\alpha_{i_1}^{qs_1} \in W_{i_1}^{(1)}$ for all $0\leq q<\frac{p_{i_1}-1}{s_1}$.
Therefore $\alpha_{i_1}^{qs_1+j} \in W_{i_1}^{(1)} \alpha_{i_1}^j$ for all $0\leq q<\frac{p_{i_1}-1}{s_1}$.
As a result, $\alpha_{i_1}^{qs_1+j} \in K$ for all $0\le q<\frac{p_{i_1}-1}{s_1}$ and all $0 \le j \le s_1-1$.
In other words, 
\begin{equation}\label{eq:u1m}
\alpha_{i_1}^t \in K \text{~for all~} t=0,1,\dots, p_{i_1}-2.
\end{equation}

Next we show that also $\alpha_{i_1}^{p_{i_1}-1} \in K$.
For every $t=1,\dots,s_1-1$ we have $0\le \lfloor \frac{p_{i_1}-1-t}{s_1} \rfloor <\frac{p_{i_1}-1}{s_1}$.
As a result,
$$
\beta^{t} \alpha_{i_1}^{t+ \lfloor \frac{p_{i_1}-1-t}{s_1} \rfloor s_1} \in W_{i_1}^{(1)}, \;
t=1,\dots,s_1-1.
$$
We obtain that, for each $t=1,\dots,s_1-1,$
$$
\beta^t \alpha_{i_1}^{p_{i_1}-1} = 
\beta^t \alpha_{i_1}^{t+ \lfloor \frac{p_{i_1}-1-t}{s_1} \rfloor s_1} 
\alpha_{i_1}^{p_{i_1}-1-t - \lfloor \frac{p_{i_1}-1-t}{s_1} \rfloor s_1}
 \in W_{i_1}^{(1)}  \alpha_{i_1}^{p_{i_1}-1-t - \lfloor \frac{p_{i_1}-1-t}{s_1} \rfloor s_1} \subseteq K.
$$
At the same time,
$$
\sum_{t=0}^{s_1 - 1}\beta^t \alpha_{i_1}^{p_{i_1}-1} \in W_{i_1}^{(2)} \subseteq K.
$$
 The last two statements together imply that
$$
\alpha_{i_1}^{p_{i_1}-1} = \sum_{t=0}^{s_1 - 1}
\beta^t \alpha_{i_1}^{p_{i_1}-1}
- \sum_{t=1}^{s_1 - 1}
\beta^t \alpha_{i_1}^{p_{i_1}-1}  \in K.
$$
Combining this with \eqref{eq:u1m}, we conclude that
$\alpha_{i_1}^t \in K$ for all $t=0,1,\dots, p_{i_1}-1$.
Recall that $1,\alpha_{i_1},\dots,\alpha_{i_1}^{p_{i_1}-1}$  is a basis of $\mathbb{F}$ over $F_{i_1}$, and that $K$ is a vector space over $F_{i_1}$, so 
$\mathbb{F} \subseteq K$.
This establishes the induction base.

Now let us fix $u_1\ge 1$ and let us assume that $\beta^{u_1'}\mathbb{F} \subseteq K$ for all $u_1'<u_1.$ To prove the induction step, we need to show that $\beta^{u_1}\mathbb{F}\subseteq K$.
Mimicking the argument that led to \eqref{eq:u1m}, we can easily show that
\begin{equation}\label{eq:ubetam}
\beta^{u_1} \alpha_{i_1}^{u_1+t} \in K
\text{~for all~} t=0,1,\dots, p_{i_1}-2.
\end{equation}
Let us show that \eqref{eq:ubetam} is also true for $t=p_{i_1}-1,$ i.e., that $\beta^{u_1} \alpha_{i_1}^{u_1+p_{i_1}-1} \in K$.
For every $1 \le t \le s_1-1-u_1$, we have $0\le \lfloor \frac{p_{i_1}-1-t}{s_1} \rfloor <\frac{p_{i_1}-1}{s_1}$.
As a result,
$$
\beta^{u_1+t} \alpha_{i_1}^{u_1+t+ \lfloor \frac{p_{i_1}-1-t}{s_1} \rfloor s_1} \in W_{i_1}^{(1)},\; t=1,\dots, s_1-1-u_1.
$$
Therefore, for all such $t$
\begin{equation}\label{eq:c1m}
\beta^{u_1+t} \alpha_{i_1}^{u_1+p_{i_1}-1} = 
\beta^{u_1+t} \alpha_{i_1}^{u_1+t+ \lfloor \frac{p_{i_1}-1-t}{s_1} \rfloor s_1}
\alpha_{i_1}^{p_{i_1}-1-t - \lfloor \frac{p_{i_1}-1-t}{s_1} \rfloor s_1} 
  \in W_{i_1}  \alpha_{i_1}^{p_{i_1}-1-t - \lfloor \frac{p_{i_1}-1-t}{s_1} \rfloor s_1} \subseteq K
\end{equation}
By the induction hypothesis, $\beta^{u_1'} \mathbb{F} \subseteq K$ for all $u_1'=0,1,\dots,u_1-1$. As a result,
\begin{equation}\label{eq:c2m}
\beta^{u_1'} \alpha_{i_1}^{u_1+p_{i_1}-1} \in K,\; u_1'=0,1,\dots,u_1-1.
\end{equation}
At the same time,
\begin{equation}\label{eq:c3m}
\sum_{t=0}^{s_1 - 1}\beta^t \alpha_{i_1}^{u_1+p_{i_1}-1}
= \Big( \sum_{t=0}^{s_1 - 1}\beta^t \alpha_{i_1}^{p_{i_1}-1} \Big) \alpha_{i_1}^{u_1}
\in W_{i_1}^{(2)} \alpha_{i_1}^{u_1} \subseteq K.
\end{equation}
Combining \eqref{eq:c1m}, \eqref{eq:c2m} and \eqref{eq:c3m}, we obtain that
$$
 \beta^{u_1} \alpha_{i_1}^{u_1+p_{i_1}-1} 
=  \sum_{t=0}^{s_1 - 1}\beta^t \alpha_{i_1}^{u_1+p_{i_1}-1}
- \sum_{u_1'=0}^{u_1-1} \beta^{u_1'} \alpha_{i_1}^{u_1+p_{i_1}-1} 
- \sum_{t=1}^{s_1-1-u_1} \beta^{u_1+t} \alpha_{i_1}^{u_1+p_{i_1}-1}
\in K.
$$
Now on account of \eqref{eq:ubetam} we can conclude that
$\beta^{u_1} \alpha_{i_1}^{u_1+t} \in K$  for all $ t =0,1,\dots, p_{i_1}-1$.
Therefore, $\beta^{u_1}\mathbb{F}\subseteq K$.
This establishes the induction step and completes the proof of the proposition.
\end{IEEEproof}

\begin{proposition}\label{prop:sumS}
For the set $S_{i_2}$ defined in \eqref{eq:defSi}, we have
$$
  \spun_{F}(S_{i_2}) + \spun_{F} (S_{i_2} \alpha_{i_2}) +\dots + 
\spun_{F} (S_{i_2} \alpha_{i_2}^{s_2-1} )=\mathbb{K}.
$$
\end{proposition}
\begin{IEEEproof}
To establish the proposition, it suffices to prove that
\begin{equation}\label{eq:wrf}
  \spun_{F}(W_{i_2}) + \spun_{F} (W_{i_2} \alpha_{i_2}) +\dots + 
\spun_{F} (W_{i_2} \alpha_{i_2}^{s_2-1} )=
\oplus_{u_2=0}^{s_2-1}\beta^{u_2 s_1} F_{i_1},
\end{equation}
where $F_{i_1}$ is defined in \eqref{eq:fi1}.
Indeed, \eqref{eq:defSi} and \eqref{eq:wrf} together imply that
\begin{align*}
 \spun_{F}(S_{i_2}) + \spun_{F} (S_{i_2} \alpha_{i_2}) +\dots + 
\spun_{F} (S_{i_2} \alpha_{i_2}^{s_2-1} ) 
 &= \oplus_{u_1=0}^{s_1-1}\oplus_{u_2=0}^{s_2-1} \oplus_{q_1=0}^{p_{i_1}-1}\beta^{u_1 + u_2 s_1} \alpha_{i_1}^{q_1} F_{i_1} \\
&= \oplus_{u=0}^{s-1} \oplus_{q_1=0}^{p_{i_1}-1}\beta^u \alpha_{i_1}^{q_1} F_{i_1}\\
& = \oplus_{u=0}^{s-1} \beta^u \mathbb{F}\\
&= \mathbb{K},
\end{align*}
where the third equality follows from the fact that the set $1,\alpha_{i_1},\dots,\alpha_{i_1}^{p_{i_1}-1}$ forms a basis of $\mathbb{F}$ over $F_{i_1}$, and the last equality follows from the fact that the set $1,\beta,\dots,\beta^{s-1}$ forms a basis of $\mathbb{K}$ over $\mathbb{F}$ (see \eqref{eq:bbk}).
Thus the proposition indeed follows from \eqref{eq:wrf}.

The proof of \eqref{eq:wrf} is exactly the same as the proof of \eqref{eq:cW} (also the same as the proof of Lemma \ref{lem1}), and therefore we do not repeat it.
\end{IEEEproof}

\begin{proposition}\label{prop:mt}
For the set $T_i$ defined in \eqref{eq:dTi}, we have
$$
  \spun_{F_{[i]}}(T_i) + \spun_{F_{[i]}} (T_i \alpha_i) +\dots + 
\spun_{F_{[i]}} (T_i \alpha_i^{s_i-1} )=\mathbb{K}.
$$
\end{proposition}
\begin{IEEEproof}
To establish the proposition, it suffices to prove that
\begin{equation}\label{eq:wts}
  \spun_{F_{[i]}}(W_i) + \spun_{F_{[i]}} (W_i \alpha_i) +\dots + 
\spun_{F_{[i]}} (W_i \alpha_i^{s_i-1} )=
\oplus_{u_i=0}^{s_i-1}\beta^{u_i t_i} F_{[i-1]},
\end{equation}
where $W_i$ is defined in \eqref{eq:dwi}, and $F_{[i-1]}$ is defined in \eqref{eq:f[i]}.
Indeed, \eqref{eq:dTi} and \eqref{eq:wts} together imply that
\begin{align*}
   \spun_{F_{[i]}}(T_i) &+ \spun_{F_{[i]}} (T_i \alpha_i) +\dots + 
\spun_{F_{[i]}} (T_i \alpha_i^{s_i-1} ) \\
 = & \oplus_{\mathbi{u}_{\sim i} \in U_{\sim i}} 
\oplus_{q_1=0}^{p_1-1}
\oplus_{q_2=0}^{p_2-1}
\dots \oplus_{q_{i-1}=0}^{p_{i-1}-1}
\Big( \beta^{\sum_{j=1}^{i-1} u_j t_j+ \sum_{j=i+1}^{h+1} u_j t_j} \prod_{1\le j<i} \alpha_j^{q_j}
\big( \oplus_{u_i=0}^{s_i-1}\beta^{u_i t_i} F_{[i-1]} \big) \Big) \\
= &  
\oplus_{u_1=0}^{s_1-1}
\oplus_{u_2=0}^{s_2-1}
\dots \oplus_{u_{h+1}=0}^{s_{h+1}-1}
\oplus_{q_1=0}^{p_1-1}
\oplus_{q_2=0}^{p_2-1}
\dots \oplus_{q_{i-1}=0}^{p_{i-1}-1}
\Big( \beta^{\sum_{j=1}^{h+1} u_j t_j} \prod_{1\le j<i} \alpha_j^{q_j} F_{[i-1]}  \Big) \\
= &  
\oplus_{u=0}^{r!-1}
\oplus_{q_1=0}^{p_1-1}
\oplus_{q_2=0}^{p_2-1}
\dots \oplus_{q_{i-1}=0}^{p_{i-1}-1}
\Big( \beta^u \prod_{1\le j<i} \alpha_j^{q_j} F_{[i-1]}  \Big) \\
 =  & \oplus_{u=0}^{r!-1} \beta^u \mathbb{F} \\
= & \mathbb{K},
\end{align*}
where the third equality follows from \eqref{eq:nsu}; the fourth equality follows from the fact that for $j=2,3,\dots,h$, the set $1,\alpha_j,\dots,\alpha_j^{p_j-1}$ forms a basis of $F_{[j-1]}$ over $F_{[j]}$ and the fact that the set $1,\alpha_1,\dots,\alpha_1^{p_1-1}$ forms a basis of $\mathbb{F}$ over $F_{[1]}$, and the last equality follows from \eqref{eq:bbk}.
Thus the proposition indeed follows from \eqref{eq:wts}.

The proof of \eqref{eq:wts} is exactly the same as the proof of \eqref{eq:cW} (also the same as the proof of Lemma \ref{lem1}), and therefore we do not repeat it.
\end{IEEEproof}

\section{Proof of Lemma \ref{lem:iSH}}\label{ap:lem:iSH}
We will prove the following more detailed claim (which implies the lemma):

\begin{claim}\label{claim2}
 For every $i\in[h]$,
\begin{equation}\label{eq:ck}
\dim_{F_{[h]}}\Big(\spun_{F_{[h]}}(S_1) + \spun_{F_{[h]}}(S_2)
+ \dots + \spun_{F_{[h]}}(S_i)\Big) = \frac{i}{d+i-k}r! \prod_{j=1}^h p_j.
\end{equation}
Moreover,  for every $i\in[h]$, there exist sets $B_i$ and $G_i$ that satisfy the following three conditions:
   \begin{enumerate}
  \item[$(i)$] $B_i$ is
a basis of $\spun_{F_{[h]}}(S_1) + \spun_{F_{[h]}}(S_2)
+ \dots + \spun_{F_{[h]}}(S_i)$ over $F_{[h]}$.
 \item[$(ii)$]
\begin{equation}\label{eq:rform}
B_i=
\bigcup_{u_{i+1} = 0}^{s_{i+1}-1}
\bigcup_{u_{i+2} = 0}^{s_{i+2}-1}
\dots
\bigcup_{u_{h+1} = 0}^{s_{h+1}-1}
\bigcup_{q_{i+1} = 0}^{p_{i+1}-1}
\bigcup_{q_{i+2} = 0}^{p_{i+2}-1}
\dots
\bigcup_{q_h = 0}^{p_h-1}
\Big( G_i \beta^{\sum_{j=i+1}^{h+1} u_j t_j} \prod_{i< j \le h} \alpha_j^{q_j} \Big).
\end{equation}
\item[$(iii)$]
\begin{equation}\label{eq:rcd}
G_i \subseteq \spun_{F_{[h]}} \Big(
\Big\{\beta^{\sum_{j=1}^i u_j t_j} \prod_{j=1}^i \alpha_j^{q_j}:
u_j=0,1,\dots,s_j-1 \text{~and~}
q_j=0,1,\dots,p_j-1
\text{~for all~} j\in[i]\Big\} \Big).
\end{equation}
\end{enumerate}
\end{claim}

{\em Proof of Claim~\ref{claim2}:}
Note that by \eqref{eq:tp} and \eqref{eq:rform},
\begin{equation}\label{eq:div}
|B_i|= \frac{r!}{t_{i+1}} \prod_{j=i+1}^h p_j |G_i|
\text{~for all~} i\in[h].
\end{equation}

We prove Claim \ref{claim2} by induction on $i$.
For $i=1$, we set $G_1=W_1$ and $B_1=S_1$, then conditions $(i)$--$(iii)$ are clearly satisfied. Moreover, it is easy to see that 
$|S_1|=\frac{1}{d+1-k}r! \prod_{j=1}^h p_j$. Together this establishes the induction base.

\vspace*{.1in}
Now let us prove the induction step. Fix $i>1$ and
 assume that the claim holds for $i-1$. By the induction hypothesis, \eqref{eq:ck} holds true, and there are a basis $B_{i-1}$ of $\spun_{F_{[h]}}(S_1) + \spun_{F_{[h]}}(S_2)
+ \dots + \spun_{F_{[h]}}(S_{i-1})$ over $F_{[h]}$ and a corresponding set $G_{i-1}$ that satisfy \eqref{eq:rform}-\eqref{eq:rcd}.
We have 
   $$
   |B_{i-1}| = \frac{i-1}{d+i-1-k}r! \prod_{j=1}^h p_j,
   $$
and so by \eqref{eq:div}
   $$
   |G_{i-1}| =\frac{i-1}{d+i-1-k} t_i \prod_{j=1}^{i-1} p_j 
=\frac{i-1}{d+i-1-k} \prod_{j=1}^{i-1} (s_j p_j).
   $$
Define the sets
\begin{align}
G_{[i]} & := \bigcup_{u_i=0}^{s_i-1} \bigcup_{q_i=0}^{p_i-1} G_{i-1} \beta^{u_i t_i} \alpha_i^{q_i},
\label{eq:dgi} \\
W_{[i]} & := \bigcup_{u_1=0}^{s_1-1}
\dots \bigcup_{u_{i-1}=0}^{s_{i-1}-1} \bigcup_{q_1=0}^{p_1-1}
\dots \bigcup_{q_{i-1}=0}^{p_{i-1}-1} \Big(W_i \beta^{\sum_{j=1}^{i-1} u_j t_j} \prod_{j=1}^{i-1} \alpha_j^{q_j}\Big). \label{eq:ww}
\end{align}
Let $G_i$ be a basis of 
$$
\spun_{F_{[h]}} (G_{[i]}) + \spun_{F_{[h]}} (W_{[i]})
$$
over $F_{[h]}$, and let $B_i$ be the set given by \eqref{eq:rform}.
It is clear that $G_i$ satisfies the condition \eqref{eq:rcd}. 

Next we show that 
$B_i$ is a basis of $\spun_{F_{[h]}}(S_1) + \spun_{F_{[h]}}(S_2) + \dots + \spun_{F_{[h]}}(S_i)$ over $F_{[h]}$.
By the induction hypothesis,
\begin{align}\label{eq:FS}
\spun_{F_{[h]}}(S_1) + &\spun_{F_{[h]}}(S_2) + \dots + \spun_{F_{[h]}}(S_{i-1})
\subseteq \spun_{F_{[h]}}(B_{i-1}) .
 \end{align}
 Now using \eqref{eq:rform}, we obtain
 \begin{align}
\spun_{F_{[h]}}(B_{i-1})&=  \spun_{F_{[h]}} \Big( 
\bigcup_{u_i = 0}^{s_i-1}
\bigcup_{u_{i+1} = 0}^{s_{i+1}-1}
\dots
\bigcup_{u_{h+1} = 0}^{s_{h+1}-1}
\bigcup_{q_i = 0}^{p_i-1}
\bigcup_{q_{i+1} = 0}^{p_{i+1}-1}
\dots
\bigcup_{q_h = 0}^{p_h-1}
\Big( G_{i-1} \beta^{\sum_{j=i}^{h+1} u_j t_j} \prod_{i\le j \le h} \alpha_j^{q_j} \Big) \Big) \nonumber \\
= & \spun_{F_{[h]}} \Big( 
\bigcup_{u_{i+1} = 0}^{s_{i+1}-1}
\bigcup_{u_{i+2} = 0}^{s_{i+2}-1}
\dots
\bigcup_{u_{h+1} = 0}^{s_{h+1}-1}
\bigcup_{q_{i+1} = 0}^{p_{i+1}-1}
\bigcup_{q_{i+2} = 0}^{p_{i+2}-1}
\dots
\bigcup_{q_h = 0}^{p_h-1}
\Big( G_{[i]} \beta^{\sum_{j=i+1}^{h+1} u_j t_j} \prod_{i< j \le h} \alpha_j^{q_j} \Big)
 \Big) \nonumber \\
\subseteq & \spun_{F_{[h]}} \Big( 
\bigcup_{u_{i+1} = 0}^{s_{i+1}-1}
\bigcup_{u_{i+2} = 0}^{s_{i+2}-1}
\dots
\bigcup_{u_{h+1} = 0}^{s_{h+1}-1}
\bigcup_{q_{i+1} = 0}^{p_{i+1}-1}
\bigcup_{q_{i+2} = 0}^{p_{i+2}-1}
\dots
\bigcup_{q_h = 0}^{p_h-1}
\Big( G_i \beta^{\sum_{j=i+1}^{h+1} u_j t_j} \prod_{i< j \le h} \alpha_j^{q_j} \Big)
 \Big) \nonumber \\
= & \spun_{F_{[h]}} ( B_i), \label{eq:eg}
\end{align}
where the second equality follows from \eqref{eq:dgi}; the inclusion on the third line follows from the definition of $G_i,$ and the last equality again follows from \eqref{eq:rform}.
According to \eqref{eq:DSi},
\begin{align}
& \spun_{F_{[h]}}(S_i) = 
\spun_{F_{[h]}} \Big(
\bigcup_{\mathbi{u}_{\sim i} \in U_{\sim i}}
\bigcup_{\mathbi{q}_{\sim i} \in Q_{\sim i}}
W_i \beta^{(\sum_{j=1;j\ne i}^{h+1} u_j t_j)} \prod_{j\in [h]\backslash \{i\}} \alpha_j^{q_j} \Big)
\nonumber \\
= & \bigcup_{u_{i+1} = 0}^{s_{i+1}-1}
\bigcup_{u_{i+2} = 0}^{s_{i+2}-1}
\dots
\bigcup_{u_{h+1} = 0}^{s_{h+1}-1}
\bigcup_{q_{i+1} = 0}^{p_{i+1}-1}
\bigcup_{q_{i+2} = 0}^{p_{i+2}-1}
\dots
\bigcup_{q_h = 0}^{p_h-1}
\Big( W_{[i]} \beta^{\sum_{j=i+1}^{h+1} u_j t_j} \prod_{i< j \le h} \alpha_j^{q_j} \Big) \nonumber \\
\subseteq & \spun_{F_{[h]}} \Big( 
\bigcup_{u_{i+1} = 0}^{s_{i+1}-1}
\bigcup_{u_{i+2} = 0}^{s_{i+2}-1}
\dots
\bigcup_{u_{h+1} = 0}^{s_{h+1}-1}
\bigcup_{q_{i+1} = 0}^{p_{i+1}-1}
\bigcup_{q_{i+2} = 0}^{p_{i+2}-1}
\dots
\bigcup_{q_h = 0}^{p_h-1}
\Big( G_i \beta^{\sum_{j=i+1}^{h+1} u_j t_j} \prod_{i< j \le h} \alpha_j^{q_j} \Big)
 \Big) \nonumber \\
= & \spun_{F_{[h]}}( B_i), \label{eq:ew}
\end{align}
where the second equality follows from \eqref{eq:ww}, and the inclusion follows from the definition of $G_i$.
Combining \eqref{eq:FS}, \eqref{eq:eg}, and \eqref{eq:ew}, we obtain that
\begin{equation}\label{eq:span}
\spun_{F_{[h]}}(S_1) + \spun_{F_{[h]}}(S_2) + \dots + \spun_{F_{[h]}}(S_i)
\subseteq \spun_{F_{[h]}}(B_i).
\end{equation}
Therefore,
$$
   |B_i| \ge \dim_{F_{[h]}}(\spun_{F_{[h]}}(S_1) + \spun_{F_{[h]}}(S_2)
+ \dots + \spun_{F_{[h]}}(S_i)).
$$
By Lemma~\ref{lem:Sh}, the number of symbols of $F_{[h]}$ downloaded from 
each of the helper nodes in order to repair
the nodes $c_1,c_2,\dots,c_i$, equals
   $
   \dim_{F_{[h]}}(\spun_{F_{[h]}}(S_1) + \spun_{F_{[h]}}(S_2)
+ \dots + \spun_{F_{[h]}}(S_i)).
     $
The cut-set bound implies that
\begin{equation}\label{eq:ge}
|B_i| \ge \dim_{F_{[h]}}(\spun_{F_{[h]}}(S_1) + \spun_{F_{[h]}}(S_2)
+ \dots + \spun_{F_{[h]}}(S_i)) \ge \frac{i}{d+i-k}r! \prod_{j=1}^h p_j.
\end{equation}

The proof of the induction step will be complete once we
show that
\begin{equation}\label{eq:le}
|B_i|\le \frac{i}{d+i-k}r! \prod_{j=1}^h p_j.
\end{equation}
Indeed, \eqref{eq:span}--\eqref{eq:le} together imply \eqref{eq:ck} and the needed fact that $B_i$ is a basis of $\spun_{F_{[h]}}(S_1) + \spun_{F_{[h]}}(S_2) + \dots + \spun_{F_{[h]}}(S_i)$ over $F_{[h]}$. 

Next let us prove \eqref{eq:le}.
From \eqref{eq:div}, this inequality will follow if we prove that
\begin{equation}\label{eq:ut}
|G_i|\le \frac{i}{d+i-k} \prod_{j=1}^i (s_jp_j).
\end{equation}
By the induction hypothesis and \eqref{eq:div}, we have
$
|G_{i-1}| = \frac{i-1}{d+i-1-k} \prod_{j=1}^{i-1} s_j p_j.
$
Combining this with \eqref{eq:dgi}--\eqref{eq:ww}, we obtain that
\begin{align*}
\left| G_{[i]} \right| & =
|G_{i-1}|s_i p_i = \frac{i-1}{d+i-1-k} \prod_{j=1}^i s_j p_j, \\
\left| W_{[i]} \right| & = |W_i|\prod_{j=1}^{i-1} s_j p_j = p_i\prod_{j=1}^{i-1} s_j p_j
= \frac{1}{d+i-k}\prod_{j=1}^i s_j p_j.
\end{align*}
Therefore,
\begin{equation}\label{eq:jh}
\begin{aligned}
|G_i| & =  \left| G_{[i]} \right|
+ \left| W_{[i]} \right| 
- \dim_{F_{[h]}} ( \spun_{F_{[h]}} (G_{[i]}) \cap
\spun_{F_{[h]}} (W_{[i]}) ) \\
& = \Big( \frac{i-1}{d+i-1-k} + \frac{1}{d+i-k} \Big) \prod_{j=1}^i (s_j p_j)
- \dim_{F_{[h]}} \Big( \spun_{F_{[h]}} (G_{[i]}) \cap
\spun_{F_{[h]}} (W_{[i]}) \Big).
\end{aligned}
\end{equation}
Since
$$
W_i \subseteq \spun_{F_{[h]}}\Big(\bigcup_{u_i=0}^{s_i-1} \bigcup_{q_i=0}^{p_i-1} \{\beta^{u_i t_i} \alpha_i^{q_i} \} \Big),
$$
we have
\begin{equation}\label{eq:cg}
G_{i-1}\odot W_i \subseteq \spun_{F_{[h]}} (G_{[i]}),
\end{equation}
where $\odot$ is defined in \eqref{eq:dod}.
According to \eqref{eq:rcd},
$$
G_{i-1} \subseteq \spun_{F_{[h]}}\Big(
\bigcup_{u_1=0}^{s_1-1}
\dots \bigcup_{u_{i-1}=0}^{s_{i-1}-1} \bigcup_{q_1=0}^{p_1-1}
\dots \bigcup_{q_{i-1}=0}^{p_{i-1}-1}  \beta^{\sum_{j=1}^{i-1} u_j t_j} \prod_{j=1}^{i-1} \alpha_j^{q_j} \Big),
$$
and consequently
$$
G_{i-1}\odot W_i \subseteq \spun_{F_{[h]}} (W_{[i]}).
$$
Combining this with \eqref{eq:cg}, we conclude that
$$
G_{i-1}\odot W_i \subseteq \spun_{F_{[h]}} (G_{[i]}) \cap
\spun_{F_{[h]}} (W_{[i]}).
$$
By the induction hypothesis, the elements in $B_{i-1}$ are linearly independent over $F_{[h]}$, and so are the elements in $G_{i-1}$. Using this together with the fact that the elements
in the set 
  $$
  \Big\{\beta^{\sum_{j=1}^i u_j t_j} \prod_{j=1}^i \alpha_j^{q_j}:
u_j=0,1,\dots,s_j-1 \text{~and~}
q_j=0,1,\dots,p_j-1
\text{~for all~} j\in[i]\Big\}
  $$
are linearly independent over $F_{[h]}$,
it is easy to see that the elements in $G_{i-1}\odot W_i$ are also linearly independent over $F_{[h]}$.
Therefore,
\begin{align*}
 \dim_{F_{[h]}} \Big( \spun_{F_{[h]}} (G_{[i]}) &\cap
\spun_{F_{[h]}} (W_{[i]}) \Big) \\
\ge & |G_{i-1}\odot W_i|= |G_{i-1}|\cdot |W_i| \\
= & \Big(\frac{i-1}{d+i-1-k} \prod_{j=1}^{i-1} (s_jp_j) \Big) p_i \\
= & \frac{i-1}{(d+i-1-k)(d+i-k)} \prod_{j=1}^i (s_jp_j) \\
= & \Big( \frac{i-1}{d+i-1-k} - \frac{i-1}{d+i-k} \Big) \prod_{j=1}^i (s_jp_j).
\end{align*}
Using this in \eqref{eq:jh}, we obtain that
\begin{align*}
|G_i| & \le \Big( \frac{i-1}{d+i-1-k} + \frac{1}{d+i-k} \Big) \prod_{j=1}^i s_j p_j
- \Big( \frac{i-1}{d+i-1-k} - \frac{i-1}{d+i-k} \Big) \prod_{j=1}^i s_jp_j  \\
& =  \frac{i}{d+i-k}  \prod_{j=1}^i s_jp_j.
\end{align*}
This establishes \eqref{eq:ut} and completes the proof of the claim. \hfill $\blacksquare$

\bibliographystyle{IEEEtranS}
\bibliography{repair}

\end{document}